\documentclass[aps,pra,twocolumn,showpacs,superscriptaddress,floatfix,10pt,nofootinbib]{revtex4-1}  
\usepackage{graphicx}
\usepackage{amssymb,amsmath,latexsym,amsthm}
\usepackage{bm}
\usepackage[dvips]{epsfig}
\usepackage{verbatim}
\usepackage{txfonts}

\renewcommand{\vec}[1]{{\bm{\mathrm{#1}}}}
\newcommand{\vhat}[1]{{\bm{\mathrm{#1}}}}
\newcommand{\ocite}[1]{Ref.~[\onlinecite{#1}]}
\let\Im\undefined
\let\Re\undefined
\DeclareMathOperator{\Tr}{Tr}
\DeclareMathOperator{\Re}{Re}
\DeclareMathOperator{\Im}{Im}
\DeclareMathOperator{\Diag}{Diag}
\newcommand{\ua}{\uparrow}
\newcommand{\da}{\downarrow}

\begin{document}

\title{Spin-orbit torques from interfacial spin-orbit coupling for various interfaces}
\author{Kyoung-Whan Kim}%
\affiliation{Institut f\"{u}r Physik, Johannes Gutenberg Universit\"{a}t Mainz, Mainz 55128, Germany}%
\affiliation{Center for Nanoscale Science and Technology, National Institute
of Standards and Technology, Gaithersburg, Maryland 20899, USA}%
\affiliation{Maryland NanoCenter, University of Maryland, College Park,
Maryland 20742, USA}%
\author{Kyung-Jin Lee}%
\affiliation{Department of Materials Science and Engineering, Korea
University, Seoul 02841, Korea}%
\affiliation{KU-KIST Graduate School of Converging Science and Technology,
Korea University, Seoul 02841, Korea}%
\author{Jairo Sinova}%
\affiliation{Institut f\"{u}r Physik, Johannes Gutenberg Universit\"{a}t Mainz, Mainz 55128, Germany}%
\affiliation{Institute of Physics, Academy of Sciences of the Czech Republic,
Cukrovarnick\'{a} 10, 162 53 Praha 6 Czech Republic}
\author{Hyun-Woo Lee}
\email{hwl@postech.ac.kr}%
\affiliation{PCTP and Department of Physics, Pohang University of Science and
Technology, Pohang 37673, Korea}%
\author{M. D. Stiles}%
\email{mark.stiles@nist.gov}%
\affiliation{Center for Nanoscale Science and Technology, National Institute
of Standards and Technology, Gaithersburg, Maryland 20899, USA}%
\date{\today}
%

\begin{abstract}
We use a perturbative approach to study the effects of interfacial spin-orbit coupling in magnetic multilayers by treating the two-dimensional Rashba model in a fully three-dimensional description of electron transport near an interface. This formalism provides a compact analytic expression for current-induced spin-orbit torques in terms of unperturbed scattering coefficients, allowing computation of spin-orbit torques for various contexts, by simply substituting scattering coefficients into the formulas. It applies to calculations of spin-orbit torques for magnetic bilayers with bulk magnetism, those with interface magnetism, a normal metal/ferromagnetic insulator junction, and a topological insulator/ferromagnet junction.  It predicts a dampinglike component of spin-orbit torque that is distinct from any intrinsic contribution or those that arise from particular spin relaxation mechanisms. We discuss the effects of proximity-induced magnetism and insertion of an additional layer and provide formulas for in-plane current, which is induced by a perpendicular bias, anisotropic magnetoresistance, and spin memory loss in the same formalism.
\end{abstract}


\maketitle

\section{Introduction\label{Sec:intro}}

Broken inversion symmetry in magnetic multilayers allows for physics that is forbidden in symmetric systems, enriching the range of their physical properties and their relevance to spintronic device applications. Spin-orbit coupling combined with inversion symmetry breaking is a core ingredient of emergent phenomena, such as the intrinsic spin Hall effect~\cite{Sinova04PRL,Inoue04PRB,Inoue06PRL,Sinova15RMP,Kato04S,Valenzuela06N}, spin-orbit torques~\cite{Miron11N,Emori13NM,Liu12S,Ryu13NN,Manchon08PRB,Matos-A09PRB,Wang12PRL,Kim12PRB,Pesin12PRB,Bijl12PRB,Kurebayashi14NN}, Dzyaloshinskii-Moriya interactions~\cite{Dzyaloshinsky58PCS,Moriya60PR,Fert80PRL,Imamura04PRB,Kim13PRL,Freimuth14JPC,Tatara16PRL}, chiral spin motive forces~\cite{Kim12PRL,Tatara13PRB}, perpendicular magnetic anisotropy~\cite{Barnes14SR,Xu12JAP,Kim16PRB}, and anisotropic magnetoresistance~\cite{Kato08PRB,Wang09EPL,Grigoryan14PRB,Zhang15PRB,Bijl12PRB,Nakayama16PRL}. The contributions from an interface are frequently modeled by a two-dimensional Rashba model~\cite{Bychkov84JETP} while those from bulk are modeled by incorporating the spin Hall effect~\cite{Sinova15RMP} into a drift-diffusion formalism~\cite{Valet93PRB} in three dimensions. Both interface and bulk contributions have the same symmetry since they originate from equivalent symmetry breaking making it difficult to distinguish mechanisms, particularly when different mechanisms are treated by different models.

Recent theoretical studies generalize the two-dimensional Rashba model in order to take into account three-dimensional transport of electrons near an interface. The two-dimensional Rashba model assumes electrons near the interface behave like a two-dimensional electron gas, thus allows only for in-plane electronic transport. Haney \emph{et al.}~\cite{Haney13PRB} generalize this model to three-dimensions by including a delta-function-like Rashba interaction at the interface and compute interfacial contributions to \emph{in-plane}-current-driven spin-orbit torques. They show that some results are qualitatively different from those of the two-dimensional model. Chen and Zhang~\cite{Chen15PRL} treat spin pumping with this model using a Green function approach. Studies of spin-orbit torques~\cite{Manchon11PRB} and anisotropic magnetoresistance~\cite{Matos-A09PRBa} in magnetic tunnel junctions under \emph{perpendicular} bias give contributions that are at least second order in the spin-orbit coupling strength due to in-plane symmetry. Refs.~\cite{Xu12JAP} and \cite{Zhang15PRB} calculate respectively magnetic anisotropy and anisotropic magnetoresistance from the three-dimensional Rashba model in particular contexts. Refs.~\cite{Amin16PRBa} and \cite{Amin16PRBb} incorporate interfacial spin-orbit coupling effects into the drift-diffusion formalism by modifying the boundary conditions. Doing so treats both interfacial and bulk spin-orbit coupling in a unified picture. Ref.~\cite{Sakanashi17arXiv} reports the solution of the drift-diffusion equation in the non-magnetic layer to capture the spin Hall effect coupled to a quantum mechanical solution in the ferromagnetic layer to capture the effects of interfacial spin-orbit coupling. 

The results of each of these theories are model specific. Studying physical consequences for a variety of systems requires recomputing them for each system, such as metallic ferromagnets in contact with heavy metals, those with insulating ferromagnets, and topological insulators in contact with a magnetic layer. Even for a single system, a work function difference between the two layers forming the interface and possible existence of proximity-induced magnetism~\cite{Ryu14NC} at the interface may complicate the analysis. General analytic expressions that are applicable for a variety of interfaces would make it easier to understand trends within systems and differences between systems.

In this paper, we develop analytic expressions for  interface contributions to current-induced spin-orbit torques by treating the interfacial spin-orbit coupling as a perturbation.  The form of the analytic expression is independent of the details of the interface [Eq.~(\ref{Eq:SOT})], written in terms of the scattering amplitudes of the interface.  All details unrelated to spin-orbit coupling are captured by those scattering amplitudes, similarly to  magnetoelectric circuit theory~\cite{Brataas00PRL,Brataas01EPJB}. This approach allows for the computation of spin-orbit torques either by  computing scattering amplitudes for a given interface or using the scattering amplitudes as fitting parameters. It is possible to compute the scattering amplitudes through first-principles calculations or by solving the Schr\"{o}dinger equation for toy models. Adopting the latter approach allows us to compute spin-orbit torques for various types of interfaces. We use the same formalism to find expressions for in-plane current, which is induced by perpendicular bias (like inverse spin Hall effect~\cite{Mosendz10PRL,Mosendz10PRB,Weiler14PRL}), anisotropic magnetoresistance (like spin Hall magnetoresistance~\cite{Huang12PRL,Nakayama13PRL,Hahn13PRB,Chen13PRB,Althammer13PRB,Kim16PRL}), and spin memory loss~\cite{Sanchez14PRL} in terms of scattering amplitudes. These are presented in Appendix~\ref{Sec(A):Appendix A}.

The three-dimensional model for interfacial spin-orbit coupling reveals effects which are absent in the two-dimensional electron gas model. In the two-dimensional model, Rashba spin-orbit coupling generates mostly fieldlike component of spin-orbit torque~\cite{Manchon08PRB,Matos-A09PRB}, while the dampinglike component becomes noticeable only when one considers an extremely resistive~\cite{Wang12PRL,Kim12PRB,Pesin12PRB,Bijl12PRB} system or a non-quadratic dispersion~\cite{Kurebayashi14NN}. In contrast, the three-dimensional model  of interfacial spin-orbit coupling reveals that in metallic magnetic bilayers a current flowing in the normal metal generates fieldlike and dampinglike components of spin-orbit torque of the same order of magnitude,
and  in some parameter regimes the dampinglike component can even be larger than the fieldlike component. 
The dampinglike contribution that we obtain here is distinct from those due to previously suggested mechanisms. For instance, an intrinsic mechanism is independent of the scattering time and vanishes for a quadratic dispersion~\cite{Inoue04PRB,Inoue06PRL}, while our result is proportional to the scattering time (thus the conductivity) and survives even for a quadratic dispersion. A detailed discussion of the distinctions is presented in Sec.~\ref{Sec:discussionA}.

Another result is a generalization of previous approaches to systems with different-Fermi-surfaces, for example, a finite exchange interaction. Previous theories~\cite{Haney13PRB,Amin16PRBa,Amin16PRBb,Zhang15PRB}, assume that all band structures are the same so that the wave vectors in the normal metal and the ferromagnet are identical, significantly simplifying the computation. However, even in the simplest model of bulk ferromagnetism, the exchange splitting $J\vec{\sigma}\cdot\vec{m}$ introduces three different wave vectors; one defined in the normal metal and one each for the majority and minority bands in the ferromagnet. In this work, we carefully take into account the different Fermi surfaces and all the resulting evanescent modes, and demonstrate that proper treatment of the evanescent modes is crucial for accurate calculation of spin-orbit torques. Indeed, they can yield significant contributions since the amplitudes of reflected states are large at the interface and their velocities are slow where the energy is close to the barrier. This makes the interaction time for these electrons quite long so they can be more strongly affected by interfacial fields. We indeed demonstrate a significant contribution to spin-orbit torque from the evanescent modes with a toy model [See Fig.~\ref{Fig:bulk magnetism result}, Eq.~(\ref{Eq:bulk magneism evanescent contribution}), and related discussions]. In the previous theories, some of effects like the anisotropic magnetoresistance~\cite{Zhang15PRB} originate from a difference between the relaxation times in majority and minority electrons in the ferromagnet, but we show that existence of the bulk magnetism by itself can also cause such effects.

We compute the current-induced contribution to the spin-orbit torque, in distinction to the electric-field-induced contribution. The former is proportional to the scattering time, thus is extrinsic. A recent paper~\cite{Kurebayashi14NN} reports the existence of intrinsic spin-orbit torque from the Berry phase, which is perpendicular to the extrinsic component. This contribution can be an explanation for dampinglike spin-orbit torque for junctions with a ferromagnetic insulator or topological insulator (See Sec.~\ref{Sec:case studyD}). We leave the calculation of intrinsic spin-orbit torque (induced by the Berry phase) in the same formalism for future work.

This paper is organized as follows. In Sec.~\ref{Sec:summary}, we summarize the core results. In Sec.~\ref{Sec:rt perturbation}, we develop a general perturbation theory of scattering matrices. First we define scattering matrices (Sec.~\ref{Sec:rt perturbation-A}) and calculate them (Sec.~\ref{Sec:rt perturbation-B}). Then we derive expressions for modified scattering matrices due to interfacial spin-orbit coupling in a perturbative regime (Sec.~\ref{Sec:rt perturbation-C}). The resulting scattering matrices allow us to write down electronic eigenstates. In Sec.~\ref{Sec:results}, we derive an expression for spin-orbit torque from these eigenstates, by calculating the angular momentum transfer (spin current) to the ferromagnet. We assume that an in-plane electrical current is applied along the $x$ direction. Since the expression is written in terms of unperturbed scattering matrices, it allows us to compute spin-orbit torque by calculating unperturbed scattering matrices for a given interface. Therefore, in Sec.~\ref{Sec:case study}, we apply our theory to various types of interfaces and various situations, such as magnetic bilayers with bulk magnetism, those with interface magnetism, a normal metal in contact with a ferromagnetic insulator, and topological insulator in contact with a metallic ferromagnet. Calculating unperturbed scattering matrices is straightforward by solving the one-dimensional Schr\"{o}dinger equation. We plot fieldlike and dampinglike components of spin-orbit torque with varying parameters and discuss the results in each subsection. In Sec.~\ref{Sec:discussion}, we make some general remarks on our theory. We compare our theory with the two-dimensional Rashba model. We also discuss how our result can be generalized when multilayer structures are considered. We discuss how proximity-induced magnetization can be considered in our theory. In Sec.~\ref{Sec:conclusion}, we summarize our results. Appendices include supplementary calculations that are not necessary for the main results.

\section{Summary of the results\label{Sec:summary}}

The purpose of this section is to summarize the behavior of the spin-orbit torques presented in Sec.~\ref{Sec:case studyA} to Sec.~\ref{Sec:case studyD}, before showing the general perturbation theory. Here we focus on the existence and relative magnitudes of spin-orbit torques generated by interfacial spin-orbit coupling saving detailed discussions for later sections. The systems under consideration are normal metal/ferromagnetic metal junctions, normal metal/ferromagnetic insulator junctions, and topological insulator/ferromagnet junctions. Throughout this paper, we refer to these as magnetic bilayers, ferromagnetic insulators, and topological insulators in short, respectively. We describe the results below and summarize them in Table~\ref{Tab:summary}.

When an in-plane current is applied to a bilayer junction, two components of spin-orbit torque can act on the magnetization $\vec{m}$. Both are perpendicular to magnetization. When a torque is odd (even) in $\vec{m}$, it is called fieldlike (dampinglike)~\cite{Garello13NN}. For instance, for a constant vector $\vec{y}$, $\vec{m}\times\vec{y}$ is fieldlike and $\vec{m}\times(\vec{y}\times\vec{m})$ is dampinglike.\footnote{These are indeed the directions of spin-orbit torque induced by an applied current along $x$. [See Eq.~(\ref{Eq:SOT})]} The names can be understood by their behaviors under time reversal: Fieldlike contributions are conservative while dampinglike contributions are dissipative. In fact, a dampinglike spin torque can act as an anti-dampinglike source, thus the terminology `dampinglike' does not mean an energy loss but originates from irreversibility.

In contrast to the two-dimensional Rashba model, we show that magnetic bilayers show both fieldlike and dampinglike components even without the Berry phase~\cite{Kurebayashi14NN} contribution and a spin relaxation mechanism~\cite{Wang12PRL,Kim12PRB,Pesin12PRB,Bijl12PRB}. The relative magnitude depends on the details of the system. We first consider a magnetic bilayer where the magnetism is dominated by an exchange splitting in the ferromagnetic bulk (not at the interface). In experiments, people usually apply a current in the normal metal side. We show that a current flowing in the normal metal ($j_{\rm N}$) generates dampinglike and fieldlike spin-orbit torques that are of the same order of magnitude. The current also flows in the ferromagnet ($j_{\rm F}$), generating a large fieldlike spin-orbit torque that can be the dominant contribution. Therefore, if $j_{\rm N}\gg j_{\rm F}$, the dampinglike and fieldlike components are on the same order of magnitude. But if $j_{\rm N}\approx j_{\rm F}$, the fieldlike component tends to dominate.

If magnetism at the interface plays a more important role than the bulk magnetism considered above, both components have similar orders of magnitude. As for the bulk magnetism case, there are two sources of spin-orbit torque, $j_{\rm N}$ and $j_{\rm F}$. We demonstrate in Sec.~\ref{Sec:case studyB} that the dampinglike contributions are mostly subtractive and the fieldlike contributions are mostly additive. Therefore, the current in the ferromagnet tends not to change the total fieldlike spin-orbit torque, but tends to reduce the dampinglike spin-orbit torque.

\begin{table}[t]
	\begin{tabular}{c|c|c|c|c}
		\hline
		System & Source & FLT & DLT & Magnitude\\
		\hline
		\hline
		2D Rashba model & $j_{\rm F}$ & $\checkmark$ & $\times$ & \\
		\hline
		Magnetic bilayer & $j_{\rm N}$ & $\checkmark$ & $\checkmark$ & FLT $\lesssim$ DLT \\
		\cline{2-5}
		(bulk magnetism) & $j_{\rm F}$ & $\checkmark$ & $\checkmark$ & FLT $\gg$ DLT \\
		\hline
		Magnetic bilayer & $j_{\rm N}$ & $\checkmark$ & $\checkmark$ & FLT $\gtrsim$ DLT \\
		\cline{2-5}
		(interface magnetism) & $j_{\rm F}$ & $\checkmark$ & $\checkmark$ & FLT $\gtrsim$ DLT \\
		\hline
		Ferromagnetic insulator & $j_{\rm N}$ & $\checkmark$ & $\times$ & \\
		\hline
		Topological insulator & $j_{\rm F}$ & $\checkmark$ & $\times$ & \\
		\hline
	\end{tabular}
	\caption{\label{Tab:summary}Behaviors of in-plane-current-induced spin-orbit torques for various systems. FLT and DLT refer to fieldlike torque and dampinglike torque respectively, and $\checkmark$ ($\times$) refers to their existence (absence). Our analytic calculation allows for the separation of contributions from a pure charge current ($j_{\rm N}$) and a pure spin current ($j_{\rm F}$) separately. A magnetic bilayer is a normal metal/ferromagnetic metal junction. Bulk magnetism originates from an exchange splitting in the ferromagnetic bulk and interface magnetism originates from a spin-dependent scattering at the interface. A ferromagnetic (topological) insulator is assumed to be attached to a normal (ferromagnetic) metal where the applied current flows. For all cases, interfacial spin-orbit coupling is present right at the interface. We present the behavior of the two-dimensional (2D) Rashba model as a reference. Since the intrinsic contribution to spin-orbit torque is not taken into account in our theory, the dampinglike component in 2D Rashba model is not considered here.}
\end{table}

For systems with a ferromagnetic insulator or a topological insulator, the dampinglike component is found to be absent.

\section{Perturbation of scattering amplitudes\label{Sec:rt perturbation}}

\subsection{Definition of the scattering matrices\label{Sec:rt perturbation-A}}

\begin{figure}
	\includegraphics[width=8.6cm]{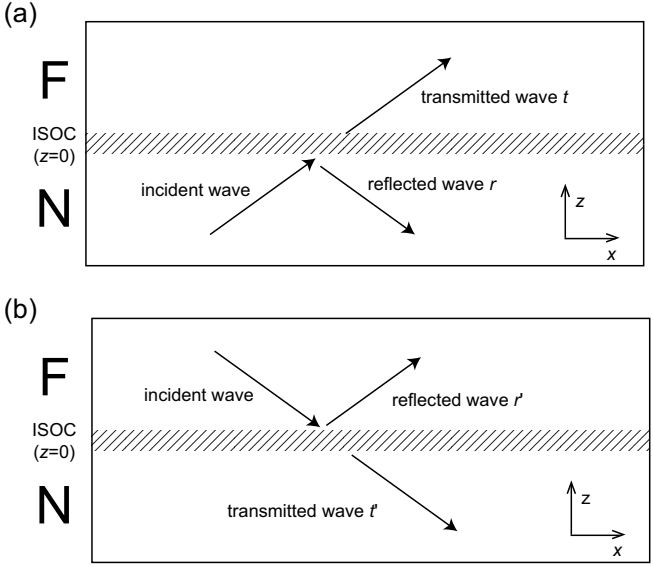}
	\caption{Scattering matrices at the interface of a
		nonmagnet($z<0$, denoted by N)/ferromagnet ($z>0$, denoted by F)
		structure. The interface is at $z=0$. $r$ and $t$
		refer to reflection and transmission matrices when the electronic state is
		incident from the normal metal layer. $r'$ and $t'$ are those when the electronic state
		is incident from the ferromagnetic layer. When the transverse mode is conserved,
		these are $2\times2$ matrices in spin space. In our model, the interfacial spin-orbit coupling (ISOC)
		is present only at $z=0$ being delta-function-like, although the thickness of
		the region drawn with finite thickness for illustration.}
	\label{Fig:scattering matrices}
\end{figure}

We consider a nonmagnet $(z<0)$/ferromagnet $(z>0)$ interface at $z=0$ where $z$ is the interface normal direction. Either materials could be insulating. We define scattering amplitudes by Fig.~\ref{Fig:scattering matrices}. The scattering of electronic states incident from the normal metal define reflection and transmission amplitudes $r$ and $t$. Those incident from the ferromagnet define $r'$ and $t'$. In the ferromagnet, there is an exchange splitting energy $J>0$. At the interface, we assume an interface potential $H_I=(\hbar^2/2m_e)\hat{\kappa}\delta(z)$, where $\hat{\kappa}$ is a $2\times2$ matrix in spin space. The delta-function-like potential describes physics on a length scale shorter than the mean free path. The effects of lattice mismatch, interface magnetism~\cite{Ryu14NC}, and interfacial spin-orbit coupling are examples. In our perturbative approach, we first ignore any interfacial spin-orbit coupling. After solving a boundary matching problem of the Schr\"{o}dinger equation at $z=0$, we obtain the scattering matrices in terms of the interface potential $\hat{\kappa}$ [Eq.~(\ref{Eq:extended matrices})]. We then add a Rashba-type interfacial spin-orbit coupling potential $\Delta\hat{\kappa}=h_R\hat{\vec{\sigma}}\cdot(\vec{k}\times\vhat{z})$ to obtain a perturbative expansion of the scattering matrices, such as $\hat{r}_\vec{k}=\hat{r}_\vec{k}^0+\Delta\hat{r}_\vec{k}$, where $\hat{r}_\vec{k}$ is the $2\times2$ reflection matrix in spin space\footnote{Throughout this paper, we denote any matrix in spin space by a symbol with a hat, $\hat{\cdot}$.} for momentum $\vec{k}$, $\hat{r}_\vec{k}^0$ is the unperturbed reflection matrix (without interfacial spin-orbit coupling), and $\Delta\hat{r}_\vec{k}$ is its correction due to interfacial spin-orbit coupling [Eq.~(\ref{Eq:scattering matrices perturbation})]. Then, the electronic eigenstates are computable analytically and play a crucial role in computing spin-orbit torque in Sec.~\ref{Sec:results}.

We now mathematically write down electronic states as definitions of scattering amplitudes illustrated in Fig.~\ref{Fig:scattering matrices}. The wave incident from the nonmagnet having the momentum $\vec{k}=(k_x,k_y,k_z)$ and spin $\sigma$ is $(1/\sqrt{V})e^{i\vec{k}\cdot\vec{r}}\xi_\sigma$ where $V$ is the volume of the system, $\vec{r}=(x,y,z)$ is the position vector, and $\xi_{\sigma}$ is the spinor for the spin $\sigma$ state. The spin quantization axis we use here is the direction of magnetization in the ferromagnet and $\sigma=\pm1$ corresponds to the minority/majority bands (with the higher/lower exchange energy). When the plane wave hits the interface at $z=0$, it scatters out. For simplicity, we assume translational symmetry over the $xy$ plane. Then, the transverse momentum is conserved, thus the scattering matrices are diagonal in transverse modes. The reflected wave has the momentum $\bar{\vec{k}}=(k_x,k_y,-k_z)$. Here we denote the reflection amplitude $r_{\vec{k}}^{\sigma'\sigma}$ by the amplitude of the scattering process $(\vec{k},\sigma)_{\rm N}\to(\bar{\vec{k}},\sigma')_{\rm N}$, where the Roman subscript/superscript N refers to the nonmagnet. The scattering state in the nonmagnet is
\begin{subequations}
\begin{equation}
\psi_{\vec{k}\sigma}^{\rm N}(z<0)=\frac{1}{\sqrt{V}}e^{i\vec{k}\cdot\vec{r}}\xi_\sigma+\frac{1}{\sqrt{V}}e^{i\bar{\vec{k}}\cdot\vec{r}}\sum_{\sigma'}r_\vec{k}^{\sigma'\sigma}\xi_{\sigma'}.
\end{equation}
The transmission matrix is defined in a similar way. Given the energy of the electronic state, the momentum $k_z$ in the ferromagnet is different from that in the nonmagnet due to the exchange splitting $J$. We denote $k_z^\sigma$ for the corresponding momentum for spin $\sigma$ band. For instance, if the kinetic energy is given by $\hbar^2|\vec{k}|^2/2m_e$ and the exchange interaction is given by $J\hat{\vec{\sigma}}\cdot\vec{m}$ where $\vec{m}$ is the unit vector along the magnetization, $\hbar^2k_z^2/2m_e=\hbar^2(k_z^+)^2/2m_e+J=\hbar^2(k_z^-)^2/2m_e-J$ defines the relation between $k_z$ and $k_z^\sigma$. Then we denote the transmission amplitude $t_\vec{k}^{\sigma'\sigma}$ by the amplitude of the scattering process $(\vec{k},\sigma)_{\rm N}\to(\vec{k}^{\sigma'},\sigma')_{\rm F}$, where the roman subscript/superscript F refers to the ferromagnet. Therefore, the scattering state in the ferromagnet is
\begin{equation}
\psi_{\vec{k}\sigma}^{\rm N}(z>0)=\frac{1}{\sqrt{V}}\sum_{\sigma'}\sqrt{\frac{|k_z|}{|k_z^{\sigma'}|}}e^{i\vec{k}^{\sigma'}\cdot\vec{r}}t_\vec{k}^{\sigma'\sigma}\xi_{\sigma'}.
\end{equation}
\end{subequations}
Here the prefactor $\sqrt{|k_z|/|k_z^{\sigma'}|}$ is introduced, in order to make the conservation of electrical charge equivalent to the unitarity of the scattering amplitudes~\cite{Datta95}. The absolute value is introduced for cases where $k_z^{\sigma'}$ is imaginary so that the transmitted wave function is evanescent. Since evanescent waves do not contribute to unitarity, this convention is arbitrary, but the choice should not affect the final expressions for physical quantities.

Now we introduce a compact matrix notation. Since the scattering amplitudes have two indices $(\sigma',\sigma)$, they are $2\times2$ matrices in spin space. We define the matrix $\hat{r}_\vec{k}=\sum_{\sigma'\sigma}\xi_{\sigma'}r_\vec{k}^{\sigma'\sigma}\xi_\sigma^\dagger$, and similarly $\hat{t}_\vec{k}$ with $t_\vec{k}^{\sigma'\sigma}$. The wave functions in this notation are
\begin{subequations}
\label{Eq:psi^N(z>0) and (z<0)}
\begin{align}
\psi_{\vec{k}\sigma}^{\rm N}(z<0)&=\frac{e^{ik_xx+ik_yy}}{\sqrt{V}}(e^{ik_zz}\hat{1}_\vec{k}+e^{-ik_zz}\hat{r}_\vec{k})\xi_\sigma,\label{Eq:psi^N(z<0)}\\
\psi_{\vec{k}\sigma}^{\rm N}(z>0)&=\frac{e^{ik_xx+ik_yy}}{\sqrt{V}}\sqrt{|k_z||\hat{K}_z|^{-1}}e^{i\hat{K}_zz}\hat{t}_\vec{k}\xi_\sigma,\label{Eq:psi^N(z>0)}
\end{align}
\end{subequations}
where $\hat{K}_z=k_z^+ u_++k_z^-u_-$ is a diagonal $2\times2$ matrix consisting of momenta in the ferromagnet for each spin band and $u_\sigma=\xi_\sigma\xi_\sigma^\dagger$ is the projection matrix to the spin $\sigma$ band. Equation~(\ref{Eq:psi^N(z>0) and (z<0)}) defines the electronic state depicted in Fig.~\ref{Fig:scattering matrices}(a).  $\hat{1}_\vec{k}$ is essentially the identity matrix, but slightly different, as we explain in the next paragraph. In a similar way, the scattering states derived from waves incident from the ferromagnet define $\hat{r}_\vec{k}'$ and $\hat{t}_\vec{k}'$ matrices in Fig.~\ref{Fig:scattering matrices}(b).
\begin{subequations}
\label{Eq:psi^F(z>0) and (z<0)}
\begin{align}
\psi_{\vec{k}\sigma}^{\rm F}(z<0)&=\frac{e^{ik_xx+ik_yy}}{\sqrt{V}}\sqrt{|k_z|^{-1}}e^{-ik_zz}\hat{t}_\vec{k}'\sqrt{|\hat{K}_z|}\xi_\sigma,\label{Eq:psi^F(z<0)}\\
\psi_{\vec{k}\sigma}^{\rm F}(z>0)&=\frac{e^{ik_xx+ik_yy}}{\sqrt{V}}\sqrt{|\hat{K}_z|^{-1}}(e^{-i\hat{K}_zz}\hat{1}_\vec{k}'+e^{i\hat{K}_zz}\hat{r}_\vec{k}')\sqrt{|\hat{K}_z|}\xi_\sigma.\label{Eq:psi^F(z>0)}
\end{align}
\end{subequations}

The matrices $\hat{1}_\vec{k}$ and $\hat{1}_\vec{l}'$ are the projection matrices \emph{to the Hilbert space}. These matrices are introduced to prevent unphysical states (not in the Hilbert space) from contributing to any physical quantities that we compute. $\hat{1}_\vec{k}$ and $\hat{1}_\vec{k}'$ are the identity matrices (the scalar 1) in the Hilbert space, but are zero, out of the Hilbert space. An electronic state is out of the Hilbert space when the incident wave is evanescent. For instance, an electronic state written as Eq.~(\ref{Eq:psi^N(z<0)}) with an imaginary $k_z$ is not in the Hilbert space in the scattering theory, so it should not contribute to any physical quantity. Thus, we define $\hat{1}_\vec{k}$ by the following projection operator:
\begin{subequations}
\begin{equation}
\hat{1}_\vec{k}=\left\{\begin{array}{cl}
1 &~\mathrm{if}~k_z~\mathrm{is~real},\\
0 &~\mathrm{if}~k_z~\mathrm{is~imaginary}.
\end{array}\right.\label{Eq:definition 1_k}
\end{equation}
In this paper, we define $\hat{r}_\vec{k}=\hat{t}_\vec{k}=0$ if the electronic state is out of the Hilbert space. Then, one can see that $\hat{r}_\vec{k}=\hat{r}_\vec{k}\hat{1}_\vec{k}$ and $\hat{t}_\vec{k}=\hat{t}_\vec{k}\hat{1}_\vec{k}$. In a similar way, $\hat{1}_\vec{k}'$ is defined by,
\begin{equation}
\hat{1}_\vec{k}'=\left\{\begin{array}{cl}
1 &~\mathrm{if~both}~k_z^\pm~\mathrm{are~real},\\
u_- &~\mathrm{if~only}~k_z^-~\mathrm{is~real},\\
0 &~\mathrm{if~both}~k_z^\pm~\mathrm{are~imaginary}.\\
\end{array}\right.\label{Eq:definition 1_k'}
\end{equation}
\end{subequations}
$\hat{1}_\vec{k}'=1$ if both majority and minority bands are propagating, $\hat{1}_\vec{k}'=u_-$ (projection to the majority band) if only majority band is propagating, and $\hat{1}_\vec{k}'=0$ if both bands are evanescent. Similarly, $\hat{r}_\vec{k}'=\hat{r}_\vec{k}'\hat{1}_\vec{k}'$ and $\hat{t}_\vec{k}'=\hat{t}_\vec{k}'\hat{1}_\vec{k}'$.

Defining the projection matrices is crucial when we consider the continuity of the wave functions at $z=0$. If the Hilbert space is not properly considered, matching Eqs.~(\ref{Eq:psi^N(z<0)}) and (\ref{Eq:psi^N(z>0)}) at $z=0$ gives $1+\hat{r}_\vec{k}=\sqrt{|k_z||\hat{K}_z|^{-1}}\hat{t}_\vec{k}$. However, it does not hold when $k_z$ is imaginary so $\hat{r}_\vec{k}=\hat{t}_\vec{k}=0$. When we project this equation to the Hilbert space by multiplying $\hat{1}_\vec{k}$, $\hat{1}_\vec{k}+\hat{r}_\vec{k}=\sqrt{|k_z||\hat{K}_z|^{-1}}\hat{t}_\vec{k}$ is the correct continuity condition. In a similar way, the continuity at $z=0$ of Eqs.~(\ref{Eq:psi^F(z<0)}) and (\ref{Eq:psi^F(z>0)}) is given by $\hat{1}_\vec{k}'+\hat{r}_\vec{k}'=\sqrt{|\hat{K}_z||k_z|^{-1}}\hat{t}_\vec{k}'$. Therefore, with the projection matrices, we can write down a single equation which holds regardless of the reality of the perpendicular momenta.

Another place where the projection matrices are crucial is the unitarity relation of the scattering amplitudes. It holds only for physical states in the Hilbert space. Therefore, the unitarity relation in our notation can be subtle. We derive the unitarity relation for the scattering amplitudes in Appendix~\ref{Sec(A):Appendix Unitary}.

\subsection{Relation to the interface potential and introduction of extended scattering matrices\label{Sec:rt perturbation-B}}

In this section, we derive explicit expressions for scattering matrices for a given interface potential. This allows defining \emph{extended} scattering matrices $(\hat{r}_{\vec{k},\rm ex},\hat{t}_{\vec{k},\rm ex},\hat{r}_{\vec{k},\rm ex}',\hat{t}_{\vec{k},\rm ex}')$ which even satisfy the continuity relation without projection. The extended scattering matrices remove the singularity of the scattering matrices,\footnote{Note that, for some momenta, the scattering matrices are zero or proportional to $u_-$, thus are not invertible matrices.} which is a main obstacle of our perturbation theory.

The explicit expressions for the scattering matrices are given by the interface potential. We start from the following interface potential at $z=0$
\begin{equation}
H_I=\frac{\hbar^2}{2m_e}\hat{\kappa}\delta(z),\label{Eq:interface potential general}
\end{equation}
where $\hat{\kappa}$ is a $2\times2$ matrix in spin space and has the dimension of the inverse of length. Solving the Schr\"{o}dinger equation gives the scattering matrices in terms of $\hat{\kappa}$. The boundary condition for the delta-function-like potential is given by the derivative mismatching condition, $\hat{\kappa}\psi_{z=0}=\partial_z\psi_{z=+0}-\partial_z\psi_{z=-0}$. After some algebra, we obtain the scattering matrices as
\begin{subequations}
	\label{Eq:extended matrices}
	\begin{align}
	\hat{t}_{\vec{k},\rm ex}&=2ik_z\sqrt{|\hat{K}_z||k_z|^{-1}}(i\hat{K}_z+ik_z-\hat{\kappa})^{-1},\\
	\hat{t}_{\vec{k},\rm ex}'&=(i\hat{K}_z+ik_z-\hat{\kappa})^{-1}2i\hat{K}_z\sqrt{|k_z||\hat{K}_z|^{-1}},\\
	\hat{r}_{\vec{k},\rm ex}&=(i\hat{K}_z+ik_z-\hat{\kappa})^{-1}(ik_z-i\hat{K}_z+\hat{\kappa}),\\
	\hat{r}_{\vec{k},\rm ex}'&=\sqrt{|\hat{K}_z|}(i\hat{K}_z+ik_z-\hat{\kappa})^{-1}(i\hat{K}_z-ik_z+\hat{\kappa})\sqrt{|\hat{K}_z|^{-1}},
	\end{align}
\end{subequations}
where we call the matrices with the subscript `ex' the extended matrices, discuss their meaning below. The expressions in Eq.~(\ref{Eq:extended matrices}) are nonzero even when the incident wave is evanescent. (For instance, $\hat{r}_\vec{k}|_{k_z=iq_z}\ne 0$.) In our convention, the scattering matrices are zero if the electronic state is evanescent because the scattering matrices capture the asymptotic behavior of the scattering process. Therefore, the scattering matrices are obtained from the extended matrices by projecting the latter to the Hilbert space.
\begin{align}
\hat{r}_\vec{k}=\hat{r}_{\vec{k},\rm ex}\hat{1}_\vec{k},~ \hat{t}_\vec{k}=\hat{t}_{\vec{k},\rm ex}\hat{1}_\vec{k},\nonumber\\ \hat{r}_\vec{k}'=\hat{r}_{\vec{k},\rm ex}'\hat{1}_\vec{k}',~ \hat{t}_\vec{k}'=\hat{t}_{\vec{k},\rm ex}'\hat{1}_\vec{k}'.
\end{align}
Now the expressions satisfy $\hat{r}_\vec{k}=\hat{r}_\vec{k}\hat{1}_\vec{k}$ and similar relations for the others.


The introduction of the extended matrices is purely mathematical. As far as physical quantities are concerned, the parts of the extended matrices out of the Hilbert space are completely arbitrary and cannot affect any physical quantity. In this paper, there are three reasons why we choose the convention in Eq.~(\ref{Eq:extended matrices}). First, it provides a natural way to write down analytic expressions valid for any momenta (even imaginary). Equation~(\ref{Eq:extended matrices}) is the result from boundary matching at $z=0$ of the Schr\"{o}dinger equation whether or not all wave vectors are real. Second, it satisfies the generalized continuity relations $1+\hat{r}_{\rm \vec{k}, ex}=\sqrt{|k_z||\hat{K}_z|^{-1}}\hat{t}_{\vec{k}, \rm ex}$ and $1+\hat{r}_{\vec{k},\rm ex}'=\sqrt{|\hat{K}_z||k_z|^{-1}}\hat{t}_{\vec{k}, \rm ex}'$ even out of the Hilbert space. This supports the idea that Eq.~(\ref{Eq:extended matrices}) is the most natural way to define the extended matrices. Third and most importantly, the extended matrices have well-defined inverses.  The singularity of $\hat{1}_\vec{k}$ and $\hat{1}_\vec{k}'$ for some momenta complicates the development of a perturbation theory, and the extended matrices give one way to resolve this difficulty.

Three remarks are in order. First, although we claim that Eq.~(\ref{Eq:extended matrices}) is the most natural form to extend out of the Hilbert space, this form depends on the normalization convention in Eqs.~(\ref{Eq:psi^N(z>0) and (z<0)}) and (\ref{Eq:psi^F(z>0) and (z<0)}) for evanescent states. But we again emphasize that the mathematical convention cannot affect calculation of physical quantities. Second, the four matrices in Eq.~(\ref{Eq:extended matrices}) are not independent since they are defined by a single matrix $\hat{\kappa}$. There are three relationships between the matrices. Two of them are the generalized continuity relations presented above. Another relationship that is derived from Eq.~(\ref{Eq:extended matrices}) is $(1+\hat{r}_{\vec{k},\rm ex})k_z^{-1}=\sqrt{|\hat{K}_z|^{-1}}(1+\hat{r}_{\vec{k},\rm ex}')\sqrt{|\hat{K}_z|}\hat{K}_z^{-1}$. Third, if $\hat{\kappa}$ is a spin-conserving Hamiltonian, $\hat{\kappa}$ and $\hat{K}_z$ commute with each other. For instance, $\sqrt{|\hat{K}|_z}$ and $\sqrt{|\hat{K}|_z^{-1}}$ in $\hat{r}_{\vec{k},\rm ex}'$ cancel so that the expression becomes simpler. The last constraint becomes simpler $(1+\hat{r}_{\vec{k},\rm ex})k_z^{-1}=(1+\hat{r}_{\vec{k},\rm ex}')\hat{K}_z^{-1}$. These features are useful for simplifying unperturbed contributions, which we consider spin conserving in the next section.

\subsection{Perturbation of scattering matrices\label{Sec:rt perturbation-C}}

To focus on the effects of interfacial spin-orbit coupling, we use a perturbative approach. Let the interface potential be
\begin{equation}
\hat{\kappa}=\hat{\kappa}_0+h_R\hat{\vec{\sigma}}\cdot(\vec{k}\times\vhat{z}),\label{Eq:kappa=kappa_0+Rashba}
\end{equation}
where the first term is the unperturbed interface potential and the second term is the interface Rashba interaction only present at $z=0$. Here $\hat{\vec{\sigma}}$ is the vector of the spin Pauli matrices, $\vhat{z}$ is the unit vector along the interface normal direction $z$, and $h_R$ is the dimensionless Rashba parameter. We treat $h_R$ perturbatively. \ocite{Haney13PRB} shows that the numerically computed spin-orbit torques are mostly linear in $h_R$, supporting this perturbative approach. We also assume that $\hat{\kappa}_0$ is spin conserving in the sense that it is diagonal in spin space. Therefore, $[\hat{\kappa}_0, u_\sigma]=[\hat{\kappa}_0, \hat{K}_z]=0$. Examples of spin-conserving potentials are spin-independent barriers and interface exchange potentials in the form of $u_m\hat{\vec{\sigma}}\cdot\vec{m}$. One interpretation of interface magnetism ($u_m$) is proximity-induced magnetism~\cite{Ryu14NC}, which is discussed in Sec.~\ref{Sec:discussionC} in more detail. The success of the conventional magnetoelectric circuit theory~\cite{Brataas00PRL,Brataas01EPJB} implies that assuming a spin conserving interface potential is reasonable.

To develop a perturbation theory, we denote unperturbed scattering matrices by a superscript 0. For instance from Eq.~(\ref{Eq:extended matrices}), $\hat{t}_{\vec{k}, \rm ex}^0=2ik_z\sqrt{|\hat{K}_z||k_z|^{-1}}(i\hat{K}_z+ik_z-\hat{\kappa}_0)^{-1}$. It is straightforward after some algebra to show that the exact scattering matrix in the presence of $h_R$ is related to the unperturbed scattering matrix as follows; $(\hat{t}_{\vec{k}, \rm ex}^0)^{-1}\hat{t}_{\vec{k}, \rm ex}=1+(h_R/2ik_z)\hat{\vec{\sigma}}\cdot(\vec{k}\times\vhat{z})(1+\hat{r}_{\vec{k},\rm ex})$.\footnote{Note that the invertibility of extended matrices is crucial for deducing this.} By multiplying $\hat{t}_{\vec{k}, \rm ex}^0$ on both sides,
\begin{equation}
\hat{t}_{\vec{k}, \rm ex}=\hat{t}_{\vec{k}, \rm ex}^0+\frac{h_R}{2ik_z}\hat{t}_{\vec{k}, \rm ex}^0\hat{\vec{\sigma}}\cdot(\vec{k}\times\vhat{z})\sqrt{|k_z||\hat{K}_z|^{-1}}\hat{t}_{\vec{k}, \rm ex},
\end{equation}
which allows a perturbative expansion with respect to $h_R$ in an iterative way. For instance, replacing $\hat{t}_{\vec{k}, \rm ex}$ in the right-hand side by $\hat{t}_{\vec{k}, \rm ex}^0$ gives the first order perturbation result for $\hat{t}_{\vec{k},\rm ex}$. From $\hat{t}_{\vec{k},\rm ex}$, the three constraints mentioned in the previous section give the rest of the extended matrices immediately. Then, projecting to the Hilbert space by multiplying by $\hat{1}_\vec{k}$ and $\hat{1}_\vec{k}'$ gives our central result for the scattering matrices.
\begin{subequations}
\label{Eq:scattering matrices perturbation}
\begin{align}
\hat{t}_{\vec{k}}&=\hat{t}_{\vec{k}}^0+\frac{h_R}{2ik_z}\hat{t}_{\vec{k}, \rm ex}^0\hat{\vec{\sigma}}\cdot(\vec{k}\times\vhat{z})(\hat{1}_\vec{k}+\hat{r}_\vec{k}^0),\\
\hat{t}_{\vec{k}}'&=\hat{t}_{\vec{k}}'^0+\frac{h_R}{2ik_z}(1+\hat{r}_{\vec{k}, \rm ex}^0)\hat{\vec{\sigma}}\cdot(\vec{k}\times\vhat{z})\hat{t}_\vec{k}'^0,\\
\hat{r}_{\vec{k}}&=\hat{r}_{\vec{k}}^0+\frac{h_R}{2ik_z}(1+\hat{r}_{\vec{k}, \rm ex}^0)\hat{\vec{\sigma}}\cdot(\vec{k}\times\vhat{z})(\hat{1}_\vec{k}+\hat{r}_\vec{k}^0),\label{Eq:result r}\\
\hat{r}_{\vec{k}}'&=\hat{r}_{\vec{k}}'^0+\frac{h_R}{2ik_z}\hat{t}_{\vec{k}, \rm ex}^0\hat{\vec{\sigma}}\cdot(\vec{k}\times\vhat{z})\hat{t}_\vec{k}'^0.
\end{align}
\end{subequations}
With Eq.~(\ref{Eq:scattering matrices perturbation}) in combination with Eqs.~(\ref{Eq:psi^N(z>0) and (z<0)}) and (\ref{Eq:psi^F(z>0) and (z<0)}), one can write down the electronic wave functions for nonzero $h_R$. Then, physical quantities can be written in terms of unperturbed scattering matrices, as we present in the next section and Appendix~\ref{Sec(A):Appendix A}. These expressions in terms of reflection and transmission coefficients can be used for general interfaces with spin-nonconserving Hamiltonians of the Rashba type.~\footnote{Even if the perturbing Hamiltonian is not in the Rashba type, our approach is still valid when one replaces $h_R\vhat{\sigma}\cdot(\vec{k}\times\vec{z})$ by the perturbing Hamiltonian.} By computing the unperturbed scattering matrices with first-principles calculations or toy models, our theory enables computing interfacial spin-orbit coupling contributions for various types of interfaces. This approach is similar to the way that one computes the spin mixing conductance~\cite{Brataas00PRL,Brataas01EPJB} in magnetoelectric circuit theory.

Three remarks are in order. First, one may notice that Eq.~(\ref{Eq:scattering matrices perturbation}) includes $1/2ik_z$ factors only, but there is no $1/2i\hat{K}_z$ factor in $\hat{r}_\vec{k}'$ and $\hat{t}_\vec{k}'$. The absence of $1/2i\hat{K}_z$ seems asymmetric since we consider all the waves incident from the normal metal and the ferromagnet. This is simply because we used the constraint $(1+\hat{r}_{\vec{k},\rm ex})^0k_z^{-1}=(1+\hat{r}_{\vec{k},\rm ex}')^0\hat{K}_z^{-1}$ to convert all $1/2i\hat{K}_z$ to $1/2ik_z$ for simplicity. Therefore, our result does not break the symmetry in the expressions. Second, in the presence of interfacial spin-orbit coupling, a bound state that does not correspond to any unperturbed state could arise. In Appendix~\ref{Sec(A):Appendix B}, we demonstrate that a bound state is not present in the perturbative regime that we consider here. Third, the presence of the extended matrices in Eq.~(\ref{Eq:scattering matrices perturbation}) is purely mathematical. This is similar to the Born approximation in scattering theory. In the Born approximation, the mathematical expression of scattering states contains virtual transitions which are not allowed due to the conservation of energy. However, such a treatment allows us to calculate the scattering states in a perturbative regime. Similarly, the presence of the extended matrices does not mean a physical transition but a mathematical artifact of the perturbative approach. Physical quantities do not depends on the extended space. The relation $(1+\hat{r}_{\vec{k},\rm ex}^0)k_z^{-1}=(1+\hat{r}_{\vec{k},\rm ex}'^0)\hat{K}_z^{-1}$ is helpful for this purpose. For instance, when we need to project $\hat{r}_{\vec{k},\rm ex}$ by $\hat{1}_\vec{k}'$ to compute a physical quantity, the relation allows expressing $\hat{r}_{\vec{k}, \rm ex}$ in terms of $\hat{r}_{\vec{k}, \rm ex}'$, so the projection by $\hat{1}_\vec{k}'$ is given by the natural relation $\hat{r}_{\vec{k}, \rm ex}'\hat{1}_\vec{k}'=\hat{r}_\vec{k}'$.

\section{Expression of Spin-orbit torque\label{Sec:results}}

\begin{figure}
\includegraphics[width=8.6cm]{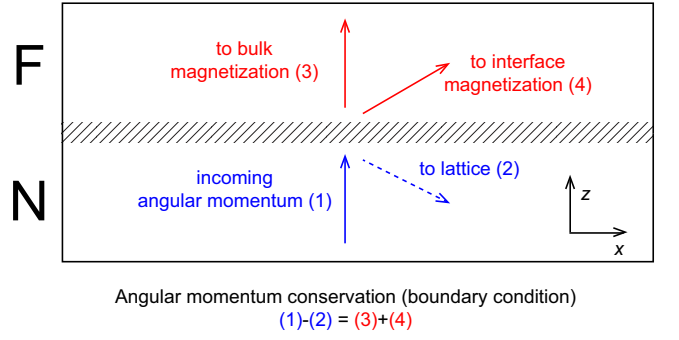}
\caption{(color online) Illustration of conservation of angular momentum at the interface.
The incoming angular momentum (1) splits into three drains; to the lattice by interfacial spin-orbit coupling (2),
to the bulk magnetism (3),
and to the interface magnetism (4).
The conservation of angular momentum implies that (1)$=$(2)$+$(3)$+$(4), which is captured by the boundary condition of the Schr\"{o}dinger equation.
Equation~(\ref{Eq:SOT}) is computed by (1)$-$(2), which is,
by the conservation of angular momentum, (3)$+$(4), the total spin-orbit torque
to magnetization.}
\label{Fig:angular momentum conservation}
\end{figure}

We consider a situation that an external current is applied. In the absence of spin-orbit coupling, angular momentum conservation suggests that the total angular momentum injected into the ferromagnet is equal to the spin current at the interface. However, in the presence of interfacial spin-orbit coupling, the spin current at $z=+0$ is not equal to the spin-orbit torque, because some of the angular momentum is transferred to the lattice. Thus, it requires a careful separation of the angular momentum flow~\cite{Amin16PRBa,Amin16PRBb} (See Fig.~\ref{Fig:angular momentum conservation}).

To develop an expression for the torque, we first ignore magnetism at the interface and restore it later. Then, the total spin-orbit torque is computed by the spin current right at the interface \emph{in the ferromagnet}, $z=+0$. For illustration, we first compute the spin current at $z=-0$ and how much angular momentum changes at the interface due to interfacial spin-orbit coupling. We compute the charge and spin current density at $z=-0$ by
\begin{align}
\hat{j}_z(\vec{r})&=-e\Tr_\vec{k}\Re[\rho\delta(\vec{r}_{\rm op}-\vec{r})v_z],\label{Eq:spin charge current}
\end{align}
where $\Re[A]=(A+A^\dagger)/2$ refers to the real part of the given matrix $A$, $\Tr_\vec{k}$ is the partial trace over $\vec{k}$ only (the result is $2\times2$ matrix in spin space), $\hat{j}_z=(j_z^e+\hat{\vec{\sigma}}\cdot\vec{j}_z^s)/2$, $\vec{j}_z^s$ is the spin current flowing along $z$ with the direction of the vector denoting the direction of spin, $j_z^e$ is the charge current along $z$, $v_z=(\hbar/m_ei)\partial_z$ is the velocity operator, $\rho=\sum_{\vec{k}\sigma'\sigma,a=\rm N/F}f_{\vec{k},\sigma',\sigma}^{\rm a}|\vec{k}\sigma';a\rangle\langle\vec{k}\sigma;a|$ is the density matrix, $f_{\vec{k}, \sigma',\sigma}^{\rm N/F}$ is the reduced density matrix for a given $\vec{k}$, $\vec{r}_{\rm op}$ is the position operator, and $\vec{r}$ is a $c$-number indicating the position at which the current density is evaluated. Here $|\vec{k}\sigma;\mathrm{N/F}\rangle$ refers to a scattering state incident from the nonmagnet (ferromagnet) that has momentum $\vec{k}$ in the nonmagnet and spin $\sigma$, that is, $\psi_{\vec{k}\sigma}^a(\vec{r})=\langle\vec{r}|\vec{k}\sigma;a\rangle$. The current is written by $\vec{j}_z(\vec{r})=(-e/2)\sum_{\vec{k}\sigma'\sigma,a=\rm N/F}f_{\vec{k}\sigma'\sigma}^{a}\langle\vec{k}\sigma;a|\{v_z,\delta(\vec{r}_{\rm op}-\vec{r})\}\hat{\vec{\sigma}}|\vec{k}\sigma';a\rangle$, and similarly for the charge current. Since we know the electronic wave functions Eqs.~(\ref{Eq:psi^N(z<0)}) and (\ref{Eq:psi^F(z<0)}), we can calculate this analytically. The delta function enables computing the matrix element without performing any integration. After some algebra,
\begin{equation}
\hat{j}_z|_{z=-0}=-\frac{eL}{hV}\sum_{\vec{k}_\perp}\int dE(\hat{f}_\vec{k}^{\rm N}\hat{1}_\vec{k}-\hat{r}_\vec{k}\hat{f}_\vec{k}^{\rm N}\hat{r}_\vec{k}^\dagger-\hat{t}_\vec{k}'\hat{f}_\vec{k}^{\rm F}\hat{t}_\vec{k}'^\dagger\hat{1}_\vec{k}).\label{Eq:current}
\end{equation}
Here, $\hat{f}_\vec{k}^{\rm N/F}=\sum_{\sigma'\sigma}\xi_{\sigma'}f_{\vec{k},\sigma',\sigma}^{\rm N/F}\xi_\sigma^\dagger$ is the matrix representation of the reduced density matrix, $L$ is the thickness of the system along $z$ direction, $h=2\pi\hbar$ is the Planck constant, the summation over $\vec{k}_\perp$ refers to the summation over all transverse momenta, and $E$ is the energy of the electron. In order to compute the contribution from $\hat{f}_\vec{k}^{\rm F}$, we assume that $\hat{f}_\vec{k}^{\rm F}$ is diagonal in $\sigma$, so that the electrons in the ferromagnet has no spin component perpendicular to the magnetization, as assumed in the magnetoelectric circuit theory. In order to convert $\sum_\vec{k}$ to $\sum_{\vec{k}_\perp}\int dE$, we use $\sum_\vec{k}=\sum_{\vec{k}_\perp}\sum_{k_z}$ and $\sum_{k_z}=(L/2\pi)\int dk_z=(m_eL/2\pi\hbar^2)\int dE/k_z$. We use $\hat{1}_\vec{k}\hat{f}_\vec{k}^{\rm N}\hat{1}=\hat{f}_\vec{k}^{\rm N}$ and  $\hat{1}_\vec{k}'\hat{f}_\vec{k}^{\rm F}\hat{1}'=\hat{f}_\vec{k}^{\rm F}$ by their definition.\footnote{There is no incident electron out of the Hilbert space.} These relations play a role in projecting the extended matrices in Eq.~(\ref{Eq:scattering matrices perturbation}) when computing physical quantities.

Equation~(\ref{Eq:current}) has the same form as the core result of the conventional magnetoelectric circuit theory~\cite{Brataas00PRL,Brataas01EPJB}. An evanescent contribution from a wave incident from the ferromagnet with $\sigma=-1$ cannot contribute to $\hat{j}_z|_{z=-0}$ (see additional $\hat{1}_\vec{k}$ factor in the last term). But in Appendix~\ref{Sec(A):Appendix A}, we show that an evanescent contribution plays an important role in an \emph{in-plane} current flow in the presence of interfacial spin-orbit coupling.

Applying an external field shifts the distribution function. In linear response regime, we approximate the Fermi surface contribution by defining chemical potentials $\Delta\hat{f}_\vec{k}^{\rm N/F}=e\Delta\hat{\mu}^{\rm N/F}\delta(E-E_F)$ where $E_F$ is the Fermi level. The delta function allows us to perform integration over $E$ in Eq.~(\ref{Eq:current}) by taking $E=E_F$. In the presence of an electrical (charge) current along $x$ direction, it shifts the electron distribution function with a finite momentum relaxation time $\tau^{\rm N}$ in the nonmagnet, $\tau^{\ua/\da}$ in the ferromagnet. Here $\ua$ and $\da$ refer to the majority ($\sigma=-1$) and minority ($\sigma=1$) bands. That is, $\Delta\hat{\mu}^{\rm N}=(E_x/m_e)\hbar k_x\tau^{\rm N}\hat{1}_\vec{k}$ and $\Delta\hat{\mu}^{\rm F}=(E_x/m_e)\hbar k_x\hat{\tau}^{\rm F}\hat{1}_\vec{k}'$ where $E_x$ is the applied electric field, and $\hat{\tau}^{\rm F}=\tau^\da u_++\tau^\ua u_-$ is a $2\times2$ matrix of the relaxation times in the ferromagnet. Then the nonequilibrium current is
\begin{align}
\hat{j}_z|_{z=-0}&=-\frac{e^2L}{hV}\sum_{\vec{k}_\perp}\left[\Delta\hat{\mu}^{\rm N}\hat{1}_\vec{k}-\hat{r}_\vec{k}\Delta\hat{\mu}^{\rm N}\hat{r}_\vec{k}^\dagger-\hat{t}_\vec{k}'\Delta\mu^{\rm F}\hat{t}_\vec{k}'^\dagger\hat{1}_\vec{k}\right]_{E=E_F}\nonumber\\
&=\frac{e^2LE_x}{2\pi m_eV}\sum_{\vec{k}_\perp}\left[k_x\hat{r}_\vec{k}\tau^N\hat{r}_\vec{k}^\dagger+k_x\hat{t}_\vec{k}'\hat{\tau}^F\hat{t}_\vec{k}'^\dagger\hat{1}_\vec{k}\right]_{E=E_F}.\label{Eq:noneq current}
\end{align}
Since the expression is given by quantities at the Fermi level, we from now on omit $[\cdots]|_{E=E_F}$ and implicitly assume that the nonequilibrium current is evaluated at the Fermi level. The simple formula Eq.~(\ref{Eq:noneq current}) gives the nonequilibrum spin and charge currents right at the interface in the nonmagnet. Without spin-orbit coupling, the system has the rotational symmetry around $z$, so $\hat{r}_\vec{k}$ is an even function of $k_x$ and $k_y$. Therefore, Eq.~(\ref{Eq:noneq current}) vanishes identically after summing up over all transverse modes. Thus, we reproduce the well-known result that there is no conventional spin-transfer torque induced by an \emph{in-plane} charge current.

However, the existence of interfacial spin-orbit coupling changes the situation drastically. Since Eq.~(\ref{Eq:scattering matrices perturbation}) includes a term which is odd in $\vec{k}$, Eq.~(\ref{Eq:noneq current}) gives rise to a finite contribution. Putting Eq.~(\ref{Eq:scattering matrices perturbation}) into Eq.~(\ref{Eq:noneq current}), we obtain
\begin{equation}
\hat{j}_z|_{z=-0}=-h_R\frac{e^2LE_x}{4\pi m_eV}\Im\sum_{\vec{k}_\perp}\frac{k_\perp^2}{k_z}(\hat{1}_\vec{k}+\hat{r}_\vec{k}^0)\hat{\sigma}_y\hat{t}_\vec{k}'^0(\hat{\tau}^{\rm F}-\tau^{\rm N})\hat{t}_\vec{k}'^{0\dagger},
\label{Eq:current at NM}
\end{equation}
where $\Im[A]=(A-A^\dagger)/2i$ refers to the imaginary part of the given matrix $A$ and $k_\perp=\sqrt{k_x^2+k_y^2}$. Here we use the unitarity relation $\hat{r}_\vec{k}^0\hat{r}_\vec{k}^{0\dagger}+\hat{t}_\vec{k}'^0\hat{t}_\vec{k}'^{0\dagger}\hat{1}_\vec{k}=\hat{1}_\vec{k}$ which is derived in Appendix~\ref{Sec(A):Appendix Unitary}. In addition, we perform an angular average of the summand: For any unit vector $\vhat{u}$, $\sum_{\vec{k}_\perp}(\vec{k}\cdot\vhat{u})\hat{\vec{\sigma}}\cdot(\vec{k}\times\vhat{z})=\sum_{\vec{k}_\perp}(k_\perp^2/2)\hat{\vec{\sigma}}\cdot(\vhat{u}\times\vhat{z})$ after taking average of contributions from all directions of $\vec{k}_\perp=k_x\vec{x}+k_y\vec{y}$ with the same magnitude. Taking $\vec{u}=\vec{x}$ yields $\hat{\sigma}_y$ in Eq.~(\ref{Eq:current at NM}).

Now we compute the discontinuity at the interface. The derivative mismatch condition $h_R\hat{\vec{\sigma}}\cdot(\vec{k}\times\vhat{z})\psi_{z=0}=\partial_z\psi_{z=+0}-\partial_z\psi_{z=-0}$, allows us to compute $\Delta\hat{j}_z\equiv\hat{j}_z|_{z=+0}-\hat{j}_z|_{z=-0}$ in terms of the wave function at $z=0$. From Eq.~(\ref{Eq:spin charge current}), $\vec{j}_z|_{z=+0}-\vec{j}_z|_{z=-0}=-h_R(e\hbar/m_e)\Tr_\vec{k}\Im[\rho\delta(\vec{r}_{\rm op}-\vec{r})\hat{\vec{\sigma}}\hat{\vec{\sigma}}\cdot(\vec{k}\times\vhat{z})]$, and similarly for the charge current. Here $\vec{r}$ is the position at the interface, so it does not have a $z$-component. Since the expression is already proportional to $h_R$, we can replace $\hat{r}_\vec{k}$ in the wave function by $\hat{r}_\vec{k}^0$. After some algebra,
\begin{equation}
\Delta\hat{j}_z=h_R\frac{e^2LE_x}{4\pi m_eV}\Im\sum_{\vec{k}_\perp}\frac{k_\perp^2}{|k_z|}\hat{\sigma}_y[(\hat{1}_\vec{k}+\hat{r}_\vec{k}^0)\tau^{\rm N}(\hat{1}_\vec{k}+\hat{r}_\vec{k}^{0\dagger})+\hat{\tau}_\vec{k}'^0\hat{\tau}^{\rm F}\hat{\tau}_\vec{k}'^{0\dagger}].\label{Eq:current discontinuity}
\end{equation}
The physical meaning of Eq.~(\ref{Eq:current discontinuity}) is the angular momentum absorption or emission at the interface due to spin-orbit coupling. Therefore, Eq.~(\ref{Eq:current discontinuity}) amounts to how much angular momentum is transferred from the lattice at the interface.

The expression for the spin-orbit torque is given by the Pauli components of $\hat{j}_z|_{z=+0}$ perpendicular to the magnetization $\vec{m}$, and $\hat{j}_z|_{z=+0}$ is given by the sum of Eqs.~(\ref{Eq:current at NM}) and (\ref{Eq:current discontinuity}). Explicitly, $\vec{T}_R=-(\hbar V/2eL)\Tr_{\sigma}[\hat{\vec{\sigma}_\perp}\hat{j}_z|_{z=+0}]$, where $\hat{\vec{\sigma}}_\perp=\hat{\vec{\sigma}}-\vec{m}(\hat{\vec{\sigma}}\cdot\vec{m})$ is the transverse part of the Pauli matrix vector to $\vec{m}$ and $\Tr_{\sigma}$ is the trace over the $2\times2$ spin space. After some algebra,
\begin{subequations}
\label{Eq:SOT}
\begin{align}
\vec{T}_R&=\Im[\mathcal{T}_R]\vec{m}\times(\vhat{y}\times\vec{m})+\Re[\mathcal{T}_R]\vec{m}\times\vhat{y},\label{Eq:TR}\\
\mathcal{T}_R&=\mathcal{T}_R^{\rm N}+\mathcal{T}_R^{\rm F},\\
\mathcal{T}_R^{\rm N}&=h_R\frac{\hbar eE_x\tau}{8\pi m_e}\sum_{\vec{k}_\perp}\frac{k_\perp^2}{k_z}(1-r_\vec{k}^{\ua*} r_\vec{k}^\da)(r_\vec{k}^\ua-r_\vec{k}^{\da*})\label{Eq:TRN}\\
\mathcal{T}_R^{\rm F}&=-h_R\frac{\hbar eE_x}{8\pi m_e}\sum_{\vec{k}_\perp}\frac{k_\perp^2}{k_z}(r_\vec{k}^\da|t_\vec{k}'^\ua|^2\tau^\ua-r_\vec{k}^{\ua*}|t_\vec{k}'^\da|^2\tau^\da)\nonumber\\
&\quad+h_R\frac{\hbar eE_x}{8\pi m_e}\sum_{\vec{k}_\perp,k_z^2<0}\frac{k_\perp^2}{|k_z|}(|t_\vec{k}'^{\ua}|^2\tau^\ua-|t_\vec{k}'^{\da}|^2\tau^\da),\label{Eq:TRF}
\end{align}
\end{subequations}
where we expanded $\hat{r}_\vec{k}^0=r_\vec{k}^\da u_++r_\vec{k}^\ua u_-$ and $\hat{t}_\vec{k}'^0=t_\vec{k}'^\da u_++t_\vec{k}'^\ua u_-$ as done in magnetoelectric circuit theory. 
Here $\ua$ is assigned for $\sigma=-1$ since $\sigma=-1$ is majority in our model. Equation~(\ref{Eq:SOT}) is the central result of this paper. The terms in Eq.~(\ref{Eq:SOT}) are the dampinglike spin-orbit torque and fieldlike spin-orbit torque respectively.

$\mathcal{T}_R^{\rm N}$ is the contribution from a current flowing in the nonmagnet. By the Drude model, the applied current is written as $n_{\rm N}e^2E_x\tau /m_e$ where $n_{\rm N}$ is the electron density in the nonmagnet. Therefore, $\tau$ multiplied by $E_x$ is proportional to the applied current. Similarly,  $\mathcal{T}_R^{\rm F}$ is the contribution from the current flowing in the ferromagnet. Especially, the second term in  $\mathcal{T}_R^{\rm F}$ is an evanescent contribution in the nonmagnet (see $k_z^2<0$). Although the wave function in the nonmagnet is evanescent, incident waves from the ferromagnet can be propagating, giving rise to a finite amount of spin-orbit torque. Such a contribution is crucial for topological insulators where the nonmagnet is insulating. We also demonstrate in Sec.~\ref{Sec:case studyA} that this contribution can also be the dominant contribution in magnetic bilayers.

In the case of torques due to the spin Hall effect in the interior of the layer, the spin Hall current proportional to $\theta_{\rm SH}E_x$ creates a spin current into the ferromagnet, where $\theta_{\rm SH}$ is the spin Hall angle. For this mechanism, the real part of the spin mixing conductance $G^{\ua\da}$ contributes to the dampinglike torque and the imaginary part contributes to the fieldlike torque~\cite{Brataas00PRL,Brataas01EPJB}. Comparing this result to Eq.~(\ref{Eq:SOT}) suggests that $\theta_{\rm SH}E_xG^{\ua\da}$ in the spin Hall effect contribution corresponds to $i\mathcal{T}_R$ in the interfacial spin-orbit coupling contribution (up to a constant factor).

Now we restore the possibility of  interface magnetism at the interface and argue that Eq.~(\ref{Eq:SOT}) is unchanged. When we add interfacial magnetism $u_m\hat{\vec{\sigma}}\cdot\vec{m}$, the boundary condition changes to $u_m\hat{\vec{\sigma}}\cdot\vec{m}\psi_{z=0}+h_R\hat{\vec{\sigma}}\cdot(\vec{k}\times\vhat{z})\psi_{z=0}=\partial_z\psi_{z=+0}-\partial_z\psi_{z=-0}$. This is nothing but the angular momentum conservation relation at the interface. The terms in the left-hand side correspond to the angular momenta transferred from the interface magnetization and the lattice. The terms in the right-hand side correspond to the incoming and outgoing angular momenta. The first term in the left-hand side and the first term in the right-hand side correspond to the (negative of) spin-transfer torque to the interfacial magnetism and the spin-transfer torque to the bulk. Therefore, the total spin-transfer torque is computed by the sum of the interfacial spin-transfer torque and the bulk spin-transfer torque, which corresponds to $\partial_z\psi_{z=+0}-u_m\hat{\vec{\sigma}}\cdot\vec{m}\psi_{z=0}$. This is the same as $\partial_z\psi_{z=-0}+h_R\hat{\vec{\sigma}}\cdot(\vec{k}\times\vhat{z})\psi_{z=0}$, which is exactly what we express in Eq.~(\ref{Eq:SOT}). Conservation  of angular momentum is summarized in Fig.~\ref{Fig:angular momentum conservation}.


\section{Spin-orbit torque for various types of interfaces\label{Sec:case study}}

In this section, we use Eq.~(\ref{Eq:SOT}) to compute spin-orbit torques for various types of interfaces. Examples includes magnetic bilayers, ferromagnetic insulators, and topological insulators as presented in Table~\ref{Tab:summary}.

\begin{figure}[b]
	\includegraphics[width=8.6cm]{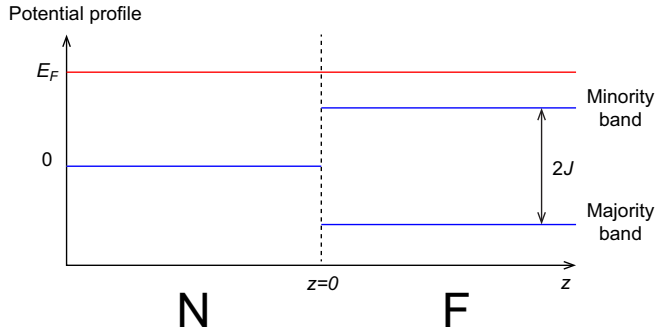}
	\caption{(color online) The potential profile (blue lines) for the model Eq.~(\ref{Eq:bulk magnetism}). The energy profile is spin-independent for $z<0$ while it has a $2J$ gap between the majority and minority bands for $z>0$. Here the red line denotes the Fermi level. The figure shows a typical situation where $E_F>J$ so that the spin polarization at the ferromagnet is incomplete, but the theory covers the whole range of positive $E_F$.}
	\label{Fig:bulk magnetism}
\end{figure}

\subsection{Magnetic bilayers - Bulk magnetism\label{Sec:case studyA}}

We start from the following unperturbed Hamiltonian.
\begin{equation}
H=-\frac{\hbar^2}{2m_e}\nabla^2+J\hat{\vec{\sigma}}\cdot\vec{m}\Theta(z),\label{Eq:bulk magnetism}
\end{equation}
where $\Theta(z)$ is the Heaviside step function representing that the bulk ferromagnetism $J$ is present only in $z>0$. The potential energy profile is presented in Fig.~\ref{Fig:bulk magnetism}. Here we assume $E_F>0$, otherwise the normal metal is insulating.

Since we have no interface potential other than spin-orbit coupling, we use Eq.~(\ref{Eq:extended matrices}) by putting $\hat{\kappa}=0$.
\begin{subequations}
\label{Eq:reflection bulk magnetism}
\begin{align}
r_\vec{k}^{\ua/\da}&=\left\{\begin{array}{cl}
                     \displaystyle\frac{k_z-k_z^\mp}{k_z+k_z^\mp} &\mathrm{if}~k_z~\mathrm{is~real,}\\
                     0 &\mathrm{if}~k_z~\mathrm{is~imaginary,}
                   \end{array}\right.\label{Eq:reflection bulk magnetism_r}\\
t_\vec{k}'^{\ua/\da}&=\left\{\begin{array}{cl}
					\displaystyle\frac{2i\sqrt{k_z^\mp|k_z|}}{ik_z+ik_z^\mp} &\mathrm{if}~k_z^\mp~\mathrm{is~real,}\\
					0 &\mathrm{if}~k_z^\mp~\mathrm{is~imaginary,}
					\end{array}\right.\label{Eq:reflection bulk magnetism_t}
\end{align}
\end{subequations}
When we define $k_F^2=2m_eE_F/\hbar^2$ and $\Delta^2=2m_eJ/\hbar^2$, each momentum has the following relations: $k_\perp^2+k_z^2=k_F^2$ and $k_\perp^2+(k_z^\pm)^2\pm \Delta^2=k_F^2$. There always are evanescent contributions from any scattered wave regardless of $E_F$ and $J$, since $k_\perp$ can be arbitrarily close to $k_F$.

The total spin-orbit torque is given by the sum of Eq.~(\ref{Eq:TRN}) and Eq.~(\ref{Eq:TRF}). 
Putting Eq.~(\ref{Eq:reflection bulk magnetism_r}) into Eq.~(\ref{Eq:TRN}) gives spin-orbit torque generated by a current flowing in the normal metal.
\begin{subequations}
\label{Eq:SOT bulk magnetism_N}
\begin{align}
\Re[\mathcal{T}_R^{\rm N}]&=-h_R\frac{\hbar eE_x\tau}{2\pi m_e}\sum_{k_\perp^2<k_F^2}k_\perp^2k_z\frac{(k_z^-)^2-|k_z^+|^2}{(k_z+k_z^-)^2|k_z+k_z^+|^2},\\
\Im[\mathcal{T}_R^{\rm N}]&=-h_R\frac{\hbar eE_x\tau}{2\pi m_e}\sum_{k_\perp^2<k_F^2}k_\perp^2k_z\frac{2k_z^-\Im[k_z^+]}{(k_z+k_z^-)^2|k_z+k_z^+|^2},
\end{align}
\end{subequations}
where $\sum_{k_\perp^2<k_F^2}$ refers to the summation over all transverse mode satisfying $k_\perp^2<k_F^2$ thus making $k_z$ real. Here, $k_z$ and $k_z^-$ are real and positive. 
The evanescent contribution $\Im[k_z^+]$ is crucial for the dampinglike component $\Im[\mathcal{T}_R^{\rm N}]$. We remark that the real and imaginary parts of Eq.~(\ref{Eq:SOT bulk magnetism_N}) have the same sign. This implies that the dampinglike and the fieldlike component of Eq.~(\ref{Eq:TRN}) in this model have the same sign.

Putting Eq.~(\ref{Eq:reflection bulk magnetism_t}) into Eq.~(\ref{Eq:TRF}) gives spin-orbit torque generated by a current flowing in the ferromagnet. Although the situation is slightly more complicated than above, the explicit expressions are similar:
\begin{subequations}
	\label{Eq:SOT bulk magnetism_F}
	\begin{align}
	\Re[\mathcal{T}_R^{\rm F}]&=-h_R\frac{\hbar eE_x\tau^\ua}{2\pi m_e}\sum_{k_\perp^2<k_F^2}k_\perp^2k_z\frac{k_z^2-|k_z^+|^2}{(k_z+k_z^-)^2|k_z+k_z^+|^2},\nonumber\\
	&\quad-h_R\frac{\hbar eE_x\tau^\da}{2\pi m_e}\sum_{k_\perp^2+\Delta^2<k_F^2}k_\perp^2k_z^+\frac{k_z^--k_z}{(k_z+k_z^+)^2(k_z+k_z^-)}\nonumber\\
	&\quad+h_R\frac{\hbar eE_x\tau^\ua}{2\pi m_e}\sum_{k_F^2<k_\perp^2<k_F^2+\Delta^2}\frac{k_\perp^2k_z^-}{\Delta^2},\\
	\Im[\mathcal{T}_R^{\rm F}]&=h_R\frac{\hbar eE_x\tau^\ua}{2\pi m_e}\sum_{k_\perp^2<k_F^2}k_\perp^2k_z^-\frac{2k_z\Im[k_z^+]}{(k_z+k_z^-)^2|k_z+k_z^+|^2}.
	\end{align}
\end{subequations}
Here, terms proportional to $\tau^{\ua}$ and $\tau^\da$ are contributions from majority and minority electron flows respectively. We remark that evanescent modes are crucial for the existence of dampinglike components. The first two terms of the real part are majority and minority counterparts of $\Re[\mathcal{T}_R^{\rm N}]$. The imaginary part has also the same form as $\Im[\mathcal{T}_R^{\rm N}]$, but only majority electrons contribute because minority electrons do not make any transition to an evanescent state in this model. The last term in the real part has no counterpart in Eq.~(\ref{Eq:SOT bulk magnetism_N}) since it originates from the imbalance between majority and minority states due to a nonzero $J$. This term originates from transition of majority electrons in the ferromagnet to evanescent states in the normal metal. We show below that this contribution is very large and can dominate the other contributions making the consideration of a finite $J$ and the resulting evanescent modes very important.

Converting the summations in Eqs.~(\ref{Eq:SOT bulk magnetism_N}) and (\ref{Eq:SOT bulk magnetism_F}) to integrations allows us to compute the spin-orbit torque as a function of $E_F/J$. To do this, we convert $\sum_{k_\perp^2<k_F^2}$ to $(A/4\pi)\int_0^{k_F^2} d(k_\perp)^2$, where $A=V/L$ is the are of the interface, and similarly for the other summations. To express all momenta in terms of $k_\perp$, we use $k_z^2=k_F^2-k_\perp^2$ and $(k_z^\pm)^2=k_F^2\mp\Delta^2-k_\perp^2$. 
There are two regimes. For $E_F\le J$, $k_z^+$ is imaginary for the whole interval of the integration $0<k_\perp^2<k_F^2$. On the other hand, for $E_F>J$, it is necessary to consider the intervals $0<k_\perp^2<k_F^2-\Delta^2$ and $k_F^2-\Delta^2<k_\perp^2<k_F^2$ separately, since $k_z^+$ is imaginary for the former and is real for the latter. Thus, it has different properties when taking the absolute value. In both cases, $k_z^-$ is always real, and $k_z$ is imaginary only when $k_F^2<k_\perp^2<k_F^2+\Delta^2$. The integration can be performed fully analytically, however, we present only numerical results due to complexity of the expressions.

\begin{figure}
	\includegraphics[width=8.6cm]{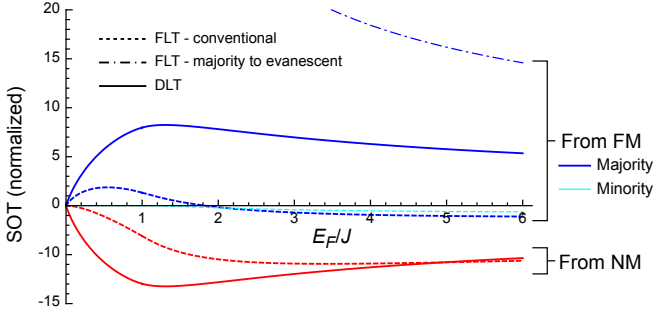}
	\caption{(color online) Spin-orbit torque (SOT) to the bulk magnetism in magnetic bilayer described by Eq.~(\ref{Eq:bulk magnetism}). The red lines are contributions from a current flowing in the normal metal, and the other lines are those from a current flowing in the ferromagnet. Among these, the blue lines are contributions from majority electrons flow and the cyan line represents contributions from minority electron flow. The dashed and dot-dashed lines are the real part, thus representing fieldlike torque (FLT), while the solid lines are the imaginary part, thus representing dampinglike torque (DLT). The dot-dashed line represents the third term in $\Re[\mathcal{T}_R^{\rm F}]$ in Eq.~(\ref{Eq:SOT bulk magnetism_F}), which originates from a nonzero value of $J$ and transitions to resulting the evanescent states. The spin-orbit torque from electrons in normal metal, majority electrons, and minority electrons are divided by $2h_R\hbar e E_x\tau Ak_F^3/\pi m_e$, $2h_R\hbar e E_x\tau^\ua Ak_F^3/\pi m_e$, and $2h_R\hbar e E_x\tau^\da Ak_F^3/\pi m_e$, respectively, and the results are dimensionless. Thus, the total spin-orbit torque is given by the weighted sum of all the contributions with the weighting factors ($\tau, \tau^\ua,\tau^\da$).
	}
	\label{Fig:bulk magnetism result}
\end{figure}

Figure~\ref{Fig:bulk magnetism result} presents (normalized) contributions of spin-orbit torques as a function of $E_F/J$. The values are divided by factors proportional to $E_x\tau$, $E_x\tau^\ua$, and $E_x\tau^\da$ for electrons in normal metal, majority electrons in the ferromagnet, and minority electrons in the ferromagnet, respectively. In most experimental situations, people apply an electrical current mainly in the normal metal. Thus we discuss the spin-orbit torque originating from a current in the normal metal first and consider the effects of a current leaking to the ferromagnet.

Red lines in Fig.~\ref{Fig:bulk magnetism result} represent fieldlike (dashed line) and dampinglike (solid line) components of spin-orbit torque induced by a current flowing in the normal metal. Unlike the two-dimensional Rashba model~\cite{Manchon08PRB,Matos-A09PRB}, the dampinglike component has the same order of magnitude as the fieldlike component and even larger for wide range of $E_F$. As we remark above, each component has the same sign. If the current leaking to the ferromagnet is sufficiently small or the ferromagnet is more resistive than the normal metal, we can consider the current to flow mainly in the normal metal. In this case, the dampinglike torque is comparable or even larger than fieldlike torque implying that experimental results for the dampinglike spin-orbit torque due to the spin Hall effect~\cite{Liu12S,Emori13NM,Ryu13NN,Liu12PRL,Zhang15NP,Nguyen16PRL} should be carefully analyzed due to the possibility of the contributions from interfacial spin-orbit coupling~\cite{Allen15PRB,Wang16PRL}.

Now we consider the contributions from the current flowing in the ferromagnet. Neglecting the unconventional term from a finite $J$ (dot-dashed line in Fig.~\ref{Fig:bulk magnetism result}), the dampinglike component (solid blue line) has the opposite sign of the normal metal contribution because the dampinglike component is generated by angular momentum carried by a spin current, which has opposite directions for these two contributions. On the other hand, since the fieldlike component originates from the current-induced spin-orbit field, the contributions from currents in both sides can act additively. Thus, the dashed blue line and the cyan line have the same sign as the dashed red line in wide range of $E_F/J$.\footnote{This argument is only valid when the propagating contributions are dominant. If $E_F$ is close to $J$, this argument is not guaranteed as shown in Fig.~\ref{Fig:bulk magnetism result} (blue and cyan dashed lines). We consider a model not containing any evanescent mode in Sec.~\ref{Sec:case studyB}, giving a clearer example of this conclusion.}

One remarkable result of this calculation is that the third term in $\Re[\mathcal{T}_R^{\rm F}]$ in Eq.~(\ref{Eq:SOT bulk magnetism_F}) is larger than the other contributions. It is around five times larger than the dampinglike components (solid lines) at $E_F\approx J$ (not shown). The origin of this term is the finite magnitude of $J$ and the resulting evanescent states. If $\tau^\ua$ has the same order of magnitude as $\tau$, this term can be the dominant contribution, illustrating the importance of accounting for the different Fermi surface. If the current flowing in the ferromagnet is at least comparable to that in the normal metal, the total spin torque is approximated by the third term in $\Re[\mathcal{T}_R^{\rm F}]$. The summation is performed in a simple analytic form:
\begin{equation}
\mathcal{T}_R\approx h_R\frac{\hbar eE_x\tau^\ua}{2\pi m_e}\sum_{k_F^2<k_\perp^2<k_F^2+\Delta^2}\frac{k_\perp^2k_z^-}{\Delta^2}=
h_R\frac{e E_x\tau^\ua A\Delta}{15\pi h}(5E_F+2J).\label{Eq:bulk magneism evanescent contribution}
\end{equation}
Since the number of electrons in the majority band remains finite when $E_F\to 0$, the contribution does not vanish in this limit, unlike the other contributions in Fig.~\ref{Fig:bulk magnetism result}.\footnote{In Fig.~\ref{Fig:bulk magnetism result}, the dot-dashed line seems divergent when $E_F\to 0$. However, this is a result of the normalizing factor $\sim k_F^3$. Figure~\ref{Fig:bulk magnetism result} does not present the dependence on $E_F$, but $E_F/J$.}

\subsection{Magnetic bilayers - Interface magnetism\label{Sec:case studyB}}

\begin{figure}
	\includegraphics[width=8.6cm]{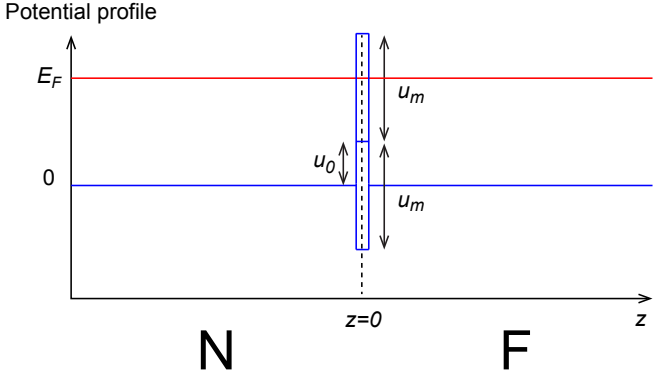}
	\caption{(color online) The potential profile (blue lines) for the model
		Eq.~(\ref{Eq:interface magnetism}). Here the spin-independent potential $u_0$
		and spin-dependent potential $u_m$ are present at $z=0$. The red line denotes the Fermi level.
		In this figure, a delta functions is represented as a square function with a finite
		height and a finite width for illustration.}
	\label{Fig:interface magnetism}
\end{figure}

We start from the following unperturbed Hamiltonian.
\begin{equation}
H=-\frac{\hbar^2}{2m_e}\nabla^2+\frac{\hbar^2}{2m_e}(u_0+u_m\hat{\vec{\sigma}}\cdot\vec{m})\delta(z),
\label{Eq:interface magnetism}
\end{equation}
where $u_0$ and $u_m$ are parameters for spin-independent and spin-dependent interface potentials. The former refers to an interface barrier and the latter refers to an interface magnetism. The interface magnetism is a possible simple model for the proximity-induced magnetism at the interface~\cite{Ryu14NC}. Since all the Fermi wave vectors are the same, there are no evanescent waves. The potential energy profile is presented in Fig.~\ref{Fig:interface magnetism}. Each side around $z=0$ is symmetric, so there is no explicit difference between the normal metal and the ferromagnet. Here we model the ferromagnet by assigning different $\tau^\ua$ and $\tau^\da$ values and assuming that angular momentum right at $z=+0$ are absorbed into the ferromagnet and make a contribution to spin-orbit torque. In other words, we implicitly assume a vanishingly small magnitude of magnetism in the ferromagnetic bulk.

We use Eq.~(\ref{Eq:extended matrices}) by putting $\hat{K}_z=k_z$ and $\hat{\kappa}=u_0+u_m\hat{\vec{\sigma}}\cdot\vec{m}$, to obtain
\begin{equation}
r_\vec{k}^{\ua\da}=\frac{u_0\mp u_m}{2ik_z-(u_0\mp u_m)},~t_\vec{k}'^{\ua\da}=1+r_\vec{k}^{\ua\da}.\label{Eq:r matrix bilayer um}
\end{equation}
Putting this into Eq.~(\ref{Eq:SOT}),
\begin{subequations}
\label{Eq:interface magnetism result}
\begin{align}
\Re[\mathcal{T}_R]&=h_R\frac{\hbar eE_x}{\pi m_e}\sum_{k_\perp^2<k_F^2} \frac{2u_0u_m(\tau+\tau_{\rm F}^e)+(u_0^2+u_m^2)\tau_{\rm F}^s}{D_\vec{k}(u_0,u_m)},
\label{Eq:interface magnetism result_re}\\
\Im[\mathcal{T}_R]&=-h_R\frac{\hbar eE_x}{\pi m_e}\sum_{k_\perp^2<k_F^2}k_z\frac{2u_0(\tau-\tau_{\rm F}^e)-2u_m\tau_{\rm F}^s}{D_\vec{k}(u_0,u_m)},
\label{Eq:interface magnetism result_im}
\end{align}
\end{subequations}
where $D_\vec{k}(u_0,u_m)=[4k_z^2+(u_0-u_m)^2][4k_z^2+(u_0+u_m)^2]/k_\perp^2 k_z$. $\tau_{\rm F}^{e/s}=(\tau^\ua\pm\tau^\da)/2$ amounts to charge/spin current flowing in the ferromagnet.

\begin{figure}
	\includegraphics[width=8.6cm]{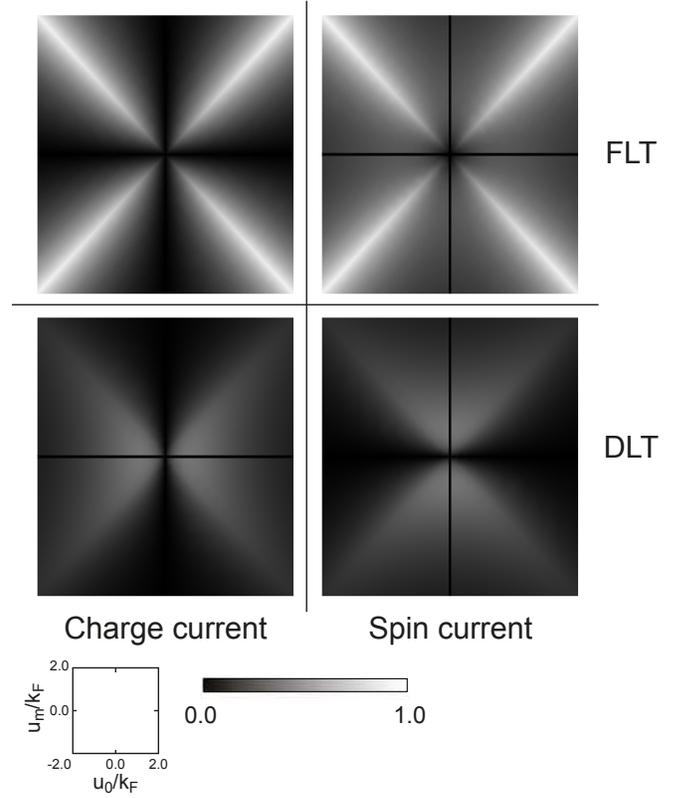}
	\caption{Magnitude of spin-orbit torque in the presence of interface magnetism. We plot spin-orbit torque as a function of $u_0/k_F$ and $u_m/k_F$. To compare magnitude clearly, we plot absolute values, discarding the signs. The upper two panels represent fieldlike components [Eq.~(\ref{Eq:interface magnetism result_re})] and the lower two panels represent dampinglike components [Eq.~(\ref{Eq:interface magnetism result_im})]. The left two panels represent charge-current-induced contributions (proportional to $\tau\pm\tau_{\rm F}^e$). The right two panels represent spin-current-induced contributions (proportional to $\tau_{\rm F}^s$). The values are divided by $h_R\hbar eE_xA/24\pi m_e k_F$ for all panels, and additionally divided by $\tau\pm\tau_{\rm F}^e$ for the charge-current-induced contributions ($+$ for FLT and $-$ for DLT) and $\tau_{\rm F}^s$ for the spin-current-induced contributions. The resulting values are dimensionless. The black lines near $u_0=0$ or $u_m=0$ are regions where spin-orbit torque is not computed due to numerical instability.
	}
	\label{Fig:interface magnetism result}
\end{figure}

We observe the following features. First, considering only charge current contributions, the fieldlike component is additive [$\tau+\tau_{\rm F}^e$ in Eq.~(\ref{Eq:interface magnetism result_re})] and the dampinglike component is subtractive [$\tau-\tau_{\rm F}^e$ in Eq.~(\ref{Eq:interface magnetism result_im})]. This observation is consistent with the discussion in Sec.~\ref{Sec:case studyA}. We can observe it here more clearly since there are no evanescent contributions in this model. Second, the charge current contributions are all zero when $u_0=0$. This means that, if an applied current is mostly flowing in the normal metal, interfacial spin-orbit torque induced by the current is proportional to the \textit{spin-independent} barrier at the interface. Third, the spin-orbit torque does not vanish even when there is no magnetism; $u_m=0$. The spin-orbit torque contribution without magnetism is attributed to our assumption that there is a vanishingly small magnetism in the ferromagnetic bulk. When angular momentum is transferred from the lattice through interfacial spin-orbit coupling, a finite amount of spin current at $z=+0$ is generated. In our approach, we assume, even when exchange in the bulk is not explicitly included, that dephasing transfers the spin angular momentum from the spin current to the bulk magnetism, giving rise to a torque. The size of the spin-orbit torque to the bulk is then determined by the conservation of angular momentum, so that the total spin torque absorbed into the ferromagnet is determined by the spin current at $z=+0$, no matter how small the bulk exchange coupling strength is in the model. In conclusion, the contributions proportional to $u_m$ are spin-orbit torque to the interface magnetism $u_m$, while the other contributions are spin-orbit torque to the ferromagnetic bulk.

To compare the relative magnitude of the fieldlike and dampinglike components, we convert the summations in Eq.~(\ref{Eq:interface magnetism result}) to integrations as in Sec.~\ref{Sec:case studyA}. We plot the absolute values of each contribution in Fig.~\ref{Fig:interface magnetism result}. Normalization factors, $\propto E_x(\tau\pm \tau_{\rm F}^e)$ or $\propto E_x\tau_{\rm F}^s$, are introduced. The total spin-orbit torque is given by the weighted sum of each panel. We observe that the fieldlike and dampinglike components are on the similar order of magnitude, but the fieldlike component is in general larger than dampinglike component in wide range of parameters. The spin-current-induced fieldlike contribution is the largest. The same model has been studied by Haney \emph{et al}.~\cite{Haney13PRB} without a perturbative approach. They show that the fieldlike component is in general larger than the dampinglike component, which is consistent with our approach.

\subsection{Ferromagnetic insulators \label{Sec:case studyC}}

\begin{figure}
\includegraphics[width=8.6cm]{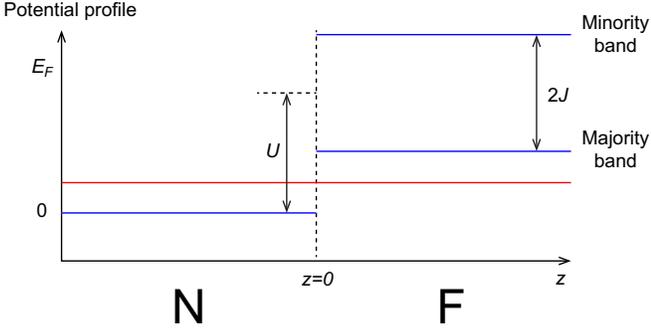}
\caption{(color online) The potential profile (blue lines) for the model Eq.~(\ref{Eq:bulk magnetism - insulator}).
In the ferromagnet, there is a spin-independent barrier $U$ that makes
the ferromagnet insulating. The Fermi level (red line) is less than
$U-J$, thus both majority and minority bands are evanescent.}
\label{Fig:bulk magnetism insulator}
\end{figure}

We start from the following unperturbed Hamiltonian.
\begin{equation}
H=-\frac{\hbar^2}{2m_e}\nabla^2+(U+J\hat{\vec{\sigma}}\cdot\vec{m})\Theta(z),
\label{Eq:bulk magnetism - insulator}
\end{equation}
where $U$ is a spin-independent potential that makes the ferromagnet ($z>0$) insulating. Thus, the Fermi level should be below $U-J$. Without loss of generality, we can assume $U>J$, otherwise there are no occupied electronic states. The potential energy profile is presented in Fig.~\ref{Fig:bulk magnetism insulator}.

Since the ferromagnet is insulating, $k_z^\pm$ are all imaginary. We define $q_z^\pm=-ik_z^\pm$, which is real and positive. The reflection amplitudes are given by the same formula as in Sec.~\ref{Sec:case studyA}.
\begin{equation}
r_\vec{k}^{\ua\da}=\frac{ik_z+q_z^\mp}{ik_z-q_z^\mp}.
\end{equation}
Since all the other momenta are imaginary, there is no contribution from $\mathcal{T}_R^{\rm F}$. Thus, we only need to compute Eq.~(\ref{Eq:TRN}):
\begin{equation}
\mathcal{T}_R=h_R\frac{\hbar^3 eE_x\tau}{2\pi m_e^2}\frac{J}{U^2-J^2}\sum_{k_\perp^2<k_F^2}k_\perp^2k_z.
\end{equation}
Since $\mathcal{T_R}$ is real, only fieldlike spin-orbit torques can survive.


We now perform the summation in the same way described in Sec.~\ref{Sec:case studyA}. After some algebra,
\begin{equation}
\mathcal{T}_R=h_R\frac{8\sqrt{2m_e}eE_x\tau A}{15\pi^2\hbar^2}\frac{JE_F^{5/2}}{U^2-J^2}.\label{Eq:spin torque bulk magnetism insulator}
\end{equation}
The spin-orbit torque vanishes at $E_F=0$ since there are no occupied states. It increases as $E_F$ increases. Equation~(\ref{Eq:spin torque bulk magnetism insulator}) has a singularity at $J=U$, but it does not  diverges at $J=U$  because as $J$ approaches  $U$, $E_F$ approaches to zero since $E_F<U-J$. The maximum of Eq.~(\ref{Eq:spin torque bulk magnetism insulator}) occurs at $E_F=U-J$; $\mathcal{T}_R\propto J(U-J)^{5/2}/(U^2-J^2)$, which has a maximum at $J=U/3$. Therefore,
\begin{equation}
\mathcal{T}_R\le h_R\frac{8\sqrt{m_e}eE_x\tau A}{45\sqrt{3}\pi^2\hbar^2}U^{3/2},
\end{equation}
which is finite.

%

\begin{figure}
\includegraphics[width=8.6cm]{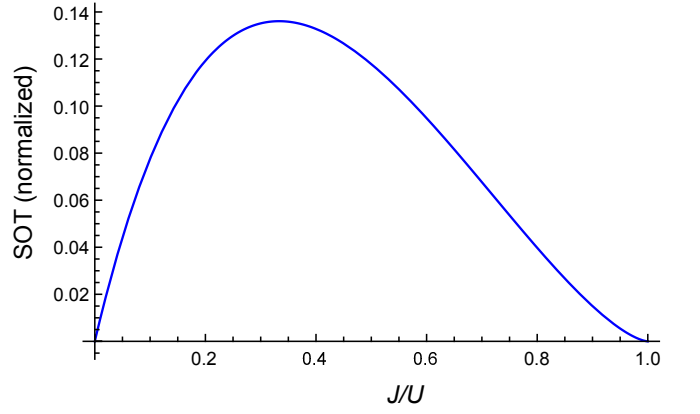}
\caption{Fieldlike spin-orbit torque in ferromagnetic insulators divided by $8h_R\sqrt{2m_e}eE_x\tau A/15\pi^2\hbar^2\times [E_F/(U-J)]^{5/2}/U^{3/2}$. The result is dimensionless.}
\label{Fig:ferro insulator result}
\end{figure}

To see the numerical behavior of $\mathcal{T}_R$, we parameterize $E_F$, $U$, and $J$ with two parameters. Since $U>J$ and $0<E_F\le U-J$, we put $J=\alpha U$ and $E_F=\beta (U-J)$ with dimensionless parameters $\alpha$ and $\beta$ satisfying $0\le \alpha,\beta\le 1$. Then,
\begin{equation}
\mathcal{T}_R=h_R\frac{8\sqrt{2m_e}eE_x\tau A}{15\pi^2\hbar^2}\times \frac{\beta^{5/2}}{U^{3/2}}\times \frac{\alpha(1-\alpha)^{3/2}}{1+\alpha}.
\end{equation}
The first factor is an overall factor proportional to the applied current. The second factor shows a simple dependence of spin torque as a function of $\beta$ and $U$. We plot the last factor in Fig.~\ref{Fig:ferro insulator result}. As we discuss above the maximum occurs at $J/U=1/3$. This plot confirms again that the spin-orbit torque is finite even when $J$ approaches $U$.

We add the interface potential $(u_0+u_m\hat{\vec{\sigma}}\cdot\vec{m})\delta(z)$ at $z=0$ and briefly see how results change. The reflection amplitudes change to
\begin{equation}
r_\vec{k}^{\ua\da}=\frac{ik_z+(q_z^\mp+u_0\mp u_m)}{ik_z-(q_z^\mp+u_0\mp u_m)}.\label{Eq:r matrix FI with um}
\end{equation}
The spin-orbit torque is then
\begin{equation}
\mathcal{T}_R=h_R\frac{\hbar eE_x\tau}{2\pi m_e}\sum_{\vec{k}_\perp}\frac{k_\perp^2k_z(q_z^-+q_z^++2u_0)(q_z^--q_z^+-2u_m)}{D_\vec{k}(u_0,u_m)},
\end{equation}
where $D_\vec{k}(u_0,u_m)=[k_z^2+(q_z^-+u_0-u_m)^2][k_z^2+(q_z^++u_0+u_m)^2]$. $\mathcal{T}_R$ is still real. It is consistent with the observation in Sec.~\ref{Sec:case studyA} that evanescent waves in the nonmagnet are crucial to get a dampinglike component. In this model, since the ferromagnet is insulating, there are no evanescent waves in the nonmagnet, thus only the fieldlike component can survive, regardless of an additional interface potential.

\subsection{Topological insulators in contact with a ferromagnet\label{Sec:case studyD}}

\begin{figure}
\includegraphics[width=8.6cm]{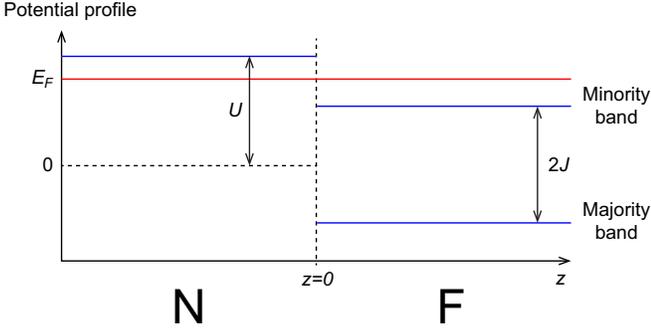}
\caption{(color online) The potential profile (blue lines) for the model
Eq.~(\ref{Eq:topological insulator}). Here the spin-independent potential $u_0$
and spin-dependent potential $u_m$ are present at $z=0$. The Fermi level (red line) is
below the barrier $U$.}
\label{Fig:topological insulator}
\end{figure}

For the case of topological insulators in contact with a ferromagnet, the nonmagnet is insulating. Thus a current flows along the ferromagnet only and the Rashba-type interaction at the interface $z=0$ gives rise to spin-orbit torque. Thus, we start from the following unperturbed Hamiltonian~\cite{Zhang09NP,Liu10PRB}.~\footnote{The model can be an oversimplication of topological surface states, but we present this model for pedagogical reasons.}
\begin{equation}
H=-\frac{\hbar^2}{2m_e}\nabla^2+U\Theta(-z)+J\hat{\vec{\sigma}}\cdot\vec{m}\Theta(z).
\label{Eq:topological insulator}
\end{equation}
Here the barrier $U$($>E_F$) makes the nonmagnetic layer insulating. Without loss of generality, we can assume that $U>J$ and $E_F>-J$, otherwise there are no occupied states. The potential profile of Eq.~(\ref{Eq:topological insulator}) is presented in Fig.~\ref{Fig:topological insulator}. In the nonmagnet, the wave vector is imaginary, so we define $q_z=-ik_z$. Then, $q_z^2+(k_z^\pm)^2=2m_e(U\mp J)/\hbar^2$. From Eq.~(\ref{Eq:extended matrices}), the transmission amplitude is given by
\begin{equation}
t_\vec{k}'^{\ua\da}=\frac{2i\sqrt{\Re[k_z^{\mp}]q_z}}{ik_z^\mp-q_z}.
\end{equation}
Since the normal metal is insulating, Eq.~(\ref{Eq:TRF}) gives spin-orbit torque. Since there is no incident wave in normal metal, $r_\vec{k}^{\ua\da}=0$, so the second term in Eq.~(\ref{Eq:TRF}) contributes only. After some algebra,
\begin{equation}
\mathcal{T}_R=h_R\frac{\hbar^3 eE_x}{4\pi m_e^2}\sum_{\vec{k}_\perp}k_\perp^2\left(\frac{\Re[k_z^-]}{U+ J}\tau^\ua-\frac{\Re[k_z^+]}{U-J}\tau^\da\right).\label{Eq:TRF_TI2}
\end{equation}
Here $k_z^-$ is real since majority waves should be propagating. But $k_z^+$ can be imaginary depending on $k_\perp$.

To perform the summation, we convert it to an integration as in Sec.~\ref{Sec:case studyA}. After some algebra,
\begin{equation}
\mathcal{T}_R=\frac{h_R\sqrt{2m_e}eE_xA}{15\pi^2 \hbar^2}\Re\left[\frac{(E_F+J)^{5/2}\tau^\ua}{U+J}-\frac{(E_F-J)^{5/2}\tau^\da}{U-J}\right].\label{Eq:TI result}
\end{equation}
Here taking the real part ($\Re$) eliminates contribution from states out of the Hilbert space. Explicitly, $\Re[(E_F-J)^{5/2}]=(E_F-J)^{5/2}\Theta(E_F-J)$, thus the second term does not contribute when $E_F<J$.

Equation~(\ref{Eq:TI result}) shows that spin-orbit torques exist even when $\tau^\ua=\tau^\da$. A similar observation is made in Appendix~\ref{Sec(A):Appendix A} that an anisotropic magnetoresistance can arise even without difference between $\tau^\ua$ and $\tau^\da$, unlike \ocite{Zhang15PRB}. This is because, in our theory, $k_F^+\ne k_F^-$. Since we break a symmetry, we obtain a torque originating from the asymmetry. As a passing remark, similarly to Sec.~\ref{Sec:case studyC}, $U-J$ in the denominator in the second term in Eq.~(\ref{Eq:TI result}) does not yield any singularity at $J=U$. This is because $E_F\le U$. When $J$ approaches to $U$, there must be a point where $J$ becomes equal or larger than $E_F$ at which the second term does not contribute.

Since Eq.~(\ref{Eq:TI result}) is real, only a fieldlike component can survive. This is not an artifact of the particular Hamiltonian that we choose [Eq.~(\ref{Eq:topological insulator})]. We remark that the second term in Eq.~(\ref{Eq:TRF}) is always real, regardless of any detail of a model. On the other hand, recent experiments~\cite{Fan14NM,Mellnik14N,Wang15PRL} report sizable dampinglike spin-orbit torques in topological insulators in contact with a ferromagnetic layer, contrary to our results. One possible cause of the dampinglike torque is that in real materials, the location of topological surface states is shifted on a nanometer scale when attached to a ferromagnet~\cite{Eremeev13PRB,Menshov13PRB,Eremeev15JMMM}. This displacement  can be a cause of a finite dampinglike torque. Another possible cause is intrinsic spin-orbit torque. In our theory, we only consider extrinsic contributions that are proportional to scattering times. However, in the two-dimensional Rashba model, the intrinsic spin-orbit torque is perpendicular to extrinsic spin-orbit torque~\cite{Kurebayashi14NN}. If similar contributions exist in our three-dimensional model, they could cause a dampinglike component.


\section{Discussion\label{Sec:discussion}}
\subsection{Comparison to the two-dimensional Rashba model\label{Sec:discussionA}}

Magnetic bilayers with bulk magnetism can behave quite differently from two-dimensional Rashba models. The two-dimensional Rashba model shows only fieldlike components~\cite{Manchon08PRB,Matos-A09PRB} unless one takes into account intrinsic spin-orbit torque from the Berry phase~\cite{Kurebayashi14NN} or a spin relaxation mechanism~\cite{Wang12PRL,Kim12PRB,Pesin12PRB,Bijl12PRB}. However, in our approach, a dampinglike spin-orbit torque with a similar order of magnitude arises if the current mostly flows in the normal metal layer. If a current flowing in the ferromagnet has a similar order of magnitude to that in the normal metal, the result shows mostly fieldlike contributions, dominated by the dot-dashed line in Fig.~\ref{Fig:bulk magnetism result}. However, the behavior is still very different from the two-dimensional model. The contribution only comes from the majority electrons in the ferromagnet and thus is proportional to $\tau^\ua$ only. On the other hand, the fieldlike spin-orbit torque derived by the two-dimensional Rashba model is proportional to the spin polarization $\propto (\tau^\ua-\tau^\da)/(\tau^\ua+\tau^\da)$. Therefore, the dominant contribution is not the counterpart of the fieldlike spin-orbit torque from the two-dimensional Rashba model.

Magnetic bilayers with interface magnetism behave similarly to the two-dimensional Rashba model. As in Fig.~\ref{Fig:interface magnetism result}, the fieldlike spin-orbit torque from spin current is the largest. In the two-dimensional Rashba model, the imbalance between the numbers of electrons in majority and minority is the primary source of spin-orbit field and the resulting fieldlike spin-orbit torque. In our model, it is modeled by $\tau^\ua\ne\tau^\da$ since we do not have a finite exchange splitting explicitly. Therefore, the upper right panel in Fig.~\ref{Fig:interface magnetism result} corresponds to the traditional contribution from the two-dimensional Rashba model. However, the spin-orbit torque contributions driven by pure charge currents $\tau_{\rm F}^e$ (left panels in Fig.~\ref{Fig:interface magnetism result}) is a unique feature of the three-dimensional model.

Systems with a ferromagnetic insulator or a topological insulator show only fieldlike spin-orbit torques. This is similar to the two-dimensional Rashba model. We observe that a current \emph{flowing at} $z=0$ (where the Rashba interaction exists) results in fieldlike spin-orbit torque while a current \emph{across} $z=0$ results in dampinglike spin-orbit torque. If one of the layers is insulating, there is no propagating wave from one to another. The situation is the same as the two-dimensional Rashba model. In the two-dimensional model, the Rashba interaction is present over the whole sample, and only in-plane electron transport is allowed. Hence, there is no possibility for electrons to cross a Rashba region, eliminating the possibility of a dampinglike contribution.

We do not consider intrinsic contributions from the Berry phase~\cite{Kurebayashi14NN}. In general, spin torques have two different contributions; extrinsic and intrinsic. The former is proportional to scattering times, while the latter is independent of scattering times. In this sense, the latter is an electric-field-induced spin torque, not a current-induced one. \ocite{Kim15PRB} highlights the subtle difference between them. The origin of an extrinsic spin torque is the change of distribution functions in the presence of an applied electric field. On the other hand, the origin of an intrinsic spin torque is the change of electronic wave functions due to an applied electric field. In the two-dimensional Rashba model, intrinsic spin torque was found to be larger than the extrinsic one in some contexts. But, it was also shown that intrinsic contributions are completely canceled out by vertex corrections~\cite{Inoue04PRB,Inoue06PRL} in metallic systems with an ideal quadratic dispersion. Therefore, the relative magnitude of the extrinsic and intrinsic spin torques depends on the situation. Similar studies would be possible in the three-dimensional model, but is beyond the scope of this paper. We defer this question for future work.

\subsection{Multilayer generalization\label{Sec:discussionB}}

The starting point of our approach is a normal metal($z<0$)/ferromagnet($z<0$) bilayer. In metallic systems consisting of layers of thicknesses larger than the mean free path, one can describe each interface separately and solve the bulk property by the spin drift-diffusion equation. Therefore, a bilayer model is sufficient to describe a multilayer system. However, if any of the layers has a thickness not much greater than the mean free path, or the system includes an insulating insertion layer at which the spin drift-diffusion equation cannot be written down, one needs to consider a multilayer situation quantum mechanically. The results of our theory will change depending on the situation. However, we here show that when we consider a normal metal ($z<0$)/any underlying structure ($0<z<L$) with a Rashba interaction at $z=0$, the reflection matrix $\hat{r}_\vec{k}$ is independent of the details of the structure underneath.

We start from the interface Hamiltonian Eq.~(\ref{Eq:interface potential general}) with $\hat{\kappa}=\hat{\kappa}_0+h_R\hat{\vec{\sigma}}\cdot(\vec{k}\times\hat{\vec{z}})$. Since we do not know any details for $z>0$, we use the transfer matrix formalism to focus on the interface at $z=0$ only. We write the wave function near $z=0$ by
\begin{subequations}
\begin{align}
\psi^0(z<0)&=\frac{e^{ik_x+ik_yy}}{\sqrt{V}}\sum_{\sigma}(e^{ik_zz}a_\sigma^R\xi_\sigma+e^{-ik_zz}a_\sigma^L\xi_\sigma),\\
\psi^0(z>0)&=\frac{e^{ik_x+ik_yy}}{\sqrt{V}}\sum_{\sigma}(e^{ik_z^\sigma z}b_\sigma^R\xi_\sigma+e^{-ik_z^\sigma z}b_\sigma^L\xi_\sigma),
\end{align}
\end{subequations}
where $R$ and $L$ refer to right-going and left-going states respectively. Here and from now on we neglect subscripts $\vec{k}\sigma$ indicating electronic states for simplicity. We define column vectors $\hat{a}^{R/L}=(a_+^{R/L},~a_-^{R/L})^T$ and $\hat{b}^{R/L}=(b_+^{R/L},~b_-^{R/L})^T$. Applying the boundary conditions at $z=0$ given by $\psi^0(+0)=\psi^0(-0)$ and $\psi^0{}'(+0)-\psi^0{}'(-0)=(2m_e\hbar^2)H_I\psi^0(0)$, we obtain the following linear relation between $\hat{b}^{R/L}$ and $\hat{a}^{R/L}$:
\begin{subequations}
\begin{gather}
\left(
\begin{array}{c}
\hat{b}^R \\
\hat{b}^L \\
\end{array}
\right)=M_0\left(
\begin{array}{c}
\hat{a}^R \\
\hat{a}^L \\
\end{array}
\right),\\
M_0^{-1}=\frac{1}{2ik_z}\left(
\begin{array}{cc}
ik_z+i\hat{K}_z-\hat{\kappa} & ik_z-i\hat{K}_z-\hat{\kappa} \\
ik_z-i\hat{K}_z+\hat{\kappa} & ik_z+i\hat{K}_z+\hat{\kappa} \\
\end{array}
\right).
\end{gather}
\end{subequations}
Here the physical meaning of $M_0$ is the transfer matrix at $z=0$ (from $z=-0$ to $z=+0$). We now express the wave function at $z=L+0$ by the matrices $\hat{c}^{R/L}$ with a suitable basis determined by the structure in $z<L$. By solving the Schr\"{o}dinger equation, we can also write down the following transfer matrix.
\begin{equation}
\left(
\begin{array}{c}
\hat{c}^R \\
\hat{c}^L \\
\end{array}
\right)=M_{0\to L}\left(
\begin{array}{c}
\hat{b}^R \\
\hat{b}^L \\
\end{array}
\right).
\end{equation}
Here $M_{0\to L}$ is the transfer matrix from $z=0$ to $z=L+0$ of which the detailed form is unnecessary here.

We consider a situation that a wave $\hat{\psi}_i$ is incident from $z<0$ and it splits up to reflected $(\hat{\psi}_r)$ and transmitted $(\hat{\psi}_t)$ parts. In the language of the transfer matrices,
\begin{equation}
\left(\begin{array}{c}
\hat{\psi}_t \\
0
\end{array}\right)
=M_{0\to L}M_0
\left(\begin{array}{c}
\hat{\psi}_i \\
\hat{\psi}_r
\end{array}\right).
\end{equation}
Inverting this,
\begin{align}
\left(\begin{array}{c}
\hat{\psi}_i \\
\hat{\psi}_r
\end{array}\right)&=\frac{1}{2ik_z}\left(
\begin{array}{cc}
ik_z+i\hat{K}_z-\hat{\kappa} & ik_z-i\hat{K}_z-\hat{\kappa} \\
ik_z-i\hat{K}_z+\hat{\kappa} & ik_z+i\hat{K}_z+\hat{\kappa} \\
\end{array}
\right)
M_{0\to L}^{-1}
\left(
\begin{array}{cc}
\hat{\psi}_t \\
0 \\
\end{array}
\right)\nonumber\\
&\equiv\frac{1}{2ik_z}\left(
\begin{array}{cc}
\hat{m}_i \\
-\hat{m}_r \\
\end{array}
\right)\hat{\psi}_t,\label{Eq:m}
\end{align}
where $\hat{m}_{i/r}$ are $2\times2$ matrices. Now the reflection matrix is given by $\hat{r}_{\rm ex}=-\hat{m}_r\hat{m}_i^{-1}$.

To expand $\hat{m}_{r/i}=\hat{m}_{r/i}^{(0)}+\hat{m}_{r/i}^{(1)}$, we split $\hat{\kappa}=\hat{\kappa}_0+\hat{\kappa}_R$. Here $\hat{\kappa}_R$ is essentially the Rashba Hamiltonian, but written in the $\hat{\cdot}$ space where the magnetization direction is along $z$. Explicitly,
\begin{equation}
\hat{\kappa}_R=h_R\left(
\begin{array}{cc}
\xi_+^\dagger \hat{\vec{\sigma}}\cdot(\vec{k}\times\vhat{z}) \xi_+ & \xi_+^\dagger \hat{\vec{\sigma}}\cdot(\vec{k}\times\vhat{z}) \xi_- \\
\xi_-^\dagger \hat{\vec{\sigma}}\cdot(\vec{k}\times\vhat{z}) \xi_+ & \xi_-^\dagger \hat{\vec{\sigma}}\cdot(\vec{k}\times\vhat{z}) \xi_-
\end{array}
\right).\label{Eq:kappa_R expression}
\end{equation}
Then,
\begin{subequations}
	\label{Eq(multi):m}
	\begin{align}
	\left(\begin{array}{c}
	\hat{m}_i^{(0)} \\
	\hat{m}_r^{(0)}
	\end{array}\right)&=
	\left(
	\begin{array}{cc}
	ik_z+i\hat{K}_z-\hat{\kappa}_0 & ik_z-i\hat{K}_z-\hat{\kappa}_0 \\
	-ik_z+i\hat{K}_z-\hat{\kappa}_0 & -ik_z-i\hat{K}_z-\hat{\kappa}_0 \\
	\end{array}
	\right)
	M_{0\to L}^{-1}
	\left(
	\begin{array}{cc}
	1 \\
	0 \\
	\end{array}
	\right),\label{Eq(multi):m0}\\
	\left(\begin{array}{c}
	\hat{m}_i^{(1)} \\
	\hat{m}_r^{(1)}
	\end{array}\right)&=-\hat{\kappa}_R
	\left(
	\begin{array}{cc}
	1 & 1 \\
	1 & 1 \\
	\end{array}
	\right)
	M_{0\to L}^{-1}
	\left(
	\begin{array}{cc}
	1 \\
	0 \\
	\end{array}
	\right).\label{Eq(multi):m1}
	\end{align}
\end{subequations}
Equation~(\ref{Eq(multi):m}) allows us to compute $\hat{m}_{r/i}^{(1)}$ in terms of unperturbed quantities. A typical way would be expressing $M_{0\to L}^{-1}(1,~0)^T$ in terms of $\hat{m}_{i/r}^{(0)}$ by inverting the matrix in front of it. Putting this into Eq.~(\ref{Eq(multi):m1}) will give $\hat{m}_{i/r}^{(1)}$ in terms of $\hat{m}_{i/r}^{(0)}$. However, the matrix inversion is very complicated. Instead, it is useful to observe from Eq.~(\ref{Eq(multi):m0}) that
\begin{equation}
\left(
\begin{array}{cc}
1 & -1 \\
1 & -1 \\
\end{array}
\right)
\left(\begin{array}{c}
\hat{m}_i^{(0)} \\
\hat{m}_r^{(0)}
\end{array}\right)=2ik_z
\left(
\begin{array}{cc}
1 & 1 \\
1 & 1 \\
\end{array}
\right)
M_{0\to d}^{-1}
\left(
\begin{array}{cc}
1 \\
0 \\
\end{array}
\right).
\end{equation}
Comparing with Eq.~(\ref{Eq(multi):m1}), we obtain
\begin{equation}
\hat{m}_i^{(1)}=\hat{m}_r^{(1)}=-\frac{1}{2ik_z}\hat{\kappa}_R(\hat{m}_i^{(0)}-\hat{m}_r^{(0)}).\label{Eq:(multi)result m}
\end{equation}
Note that this expression is not perturbative since we have not assumed a small $h_R$ up to this point.

From Eq.~(\ref{Eq:(multi)result m}), we calculate $\hat{r}_{\rm ex}=-\hat{m}_r\hat{m}_i^{-1}$. First, we perturbatively expand $\hat{r}_{\rm ex}$ by $\hat{r}_{\rm ex}=\hat{r}_{\rm ex}^{(0)}+\hat{r}_{\rm ex}^{(1)}+\hat{r}_{\rm ex}^{(2)}\cdots$, where $\hat{r}_{\rm ex}^{(n)}$ is the $n$-th order Rashba contribution. After some algebra, we obtain
\begin{subequations}
\begin{align}
\hat{r}_{\rm ex}^{(n)}&=2ik_z\hat{G}(\hat{\kappa}_R\hat{G})^n,\label{Eq:multi generalization of r}\\
\hat{G}&=\frac{1}{2ik_z}(1+\hat{r}_{\rm ex}^{(0)}).
\end{align}
\end{subequations}
Here we used $\hat{m}_i^{-1}=(\tilde{m}_i^{(0)})^{-1}-(\tilde{m}_i^{(0)})^{-1}\tilde{m}_i^{(1)}(\tilde{m}_i^{(0)})^{-1}+(\tilde{m}_i^{(0)})^{-1}\tilde{m}_i^{(1)}(\tilde{m}_i^{(0)})^{-1}\tilde{m}_i^{(1)}(\tilde{m}_i^{(0)})^{-1}+\cdots$ and $\hat{\kappa}_R\hat{G}=-\hat{m}_i^{(1)}(\hat{m}_i^{(0)})^{-1}$ actively. Taking $n=1$ gives the same result as Eq.~(\ref{Eq:result r}). Since we do not assume anything about the underlying structure in $z>0$, our result on the reflection matrix holds for arbitrary underlying structures.

Three remarks are in order. First, although the same expression holds only for the reflection matrix, it is very useful for some situations. If one looks into a response of the normal metal induced by a current flow in the normal metal,
the expression only includes $\hat{r}_\vec{k}$. The anisotropic magnetoresistance calculated in Appendix~\ref{Sec(A):Appendix A} is an example. Second, this derivation is a mathematical result, so the results holds in the extended space. Projecting Eq.~(\ref{Eq:multi generalization of r}) by $\hat{1}$ does indeed give Eq.~(\ref{Eq:result r}). Third, the derivation by the transfer matrix is somewhat more abstract than the scattering formalism in Sec.~\ref{Sec:rt perturbation}, but it allows easily generalizing our result up to any higher order contributions from $h_R$. The second order term is used in Appendix~\ref{Sec(A):Appendix A}.

\subsection{Effects of proximity-induced magnetism\label{Sec:discussionC}}

We model proximity-induced magnetism as magnetism right at the interface in Sec.~\ref{Sec:case studyB} and \ref{Sec:case studyC}. In this section, we present how one can treat effects of interface magnetism more generally.

We first consider a situation without interface magnetism, and then treat $u_m$ separately. Let $\hat{r}_{\rm ex}^{u_m=0}$ be the reflection matrix in the absence of interface magnetism. Then, it would be valuable to see how the scattering coefficients change in the presence of $u_m$. The transfer matrix formalism in Sec.~\ref{Sec:discussionB} allows calculating the contributions from $u_m$  perturbatively. When we replace $\hat{\kappa}_R$ in Eq.~(\ref{Eq:multi generalization of r}) by $u_m\hat{\sigma}_z$, we obtain $\hat{r}_{\rm ex}=\hat{r}_{\rm ex}^{u_m=0}+\hat{r}_{\rm ex}^{(1)}+\hat{r}_{\rm ex}^{(2)}+\cdots$ where
$\hat{r}_{\rm ex}^{(n)}=-2ik_z\hat{G}(u_m\hat{\sigma}_z\hat{G})^{n}$ and $\hat{G}=(-1/2ik_z)(1+\hat{r}_{\rm ex}^{u_m=0})$. Here we use the fact that any two diagonal matrices commute with each other. The result is given by the sum of a geometric series. After some algebra,
\begin{equation}
\hat{r}_{\rm ex}=\hat{r}_{\rm ex}^{u_m=0}-\frac{1+\hat{r}_{\rm ex}^{u_m=0}}{2ik_zu_m^{-1}\hat{\sigma}_z(1+\hat{r}_{\rm ex}^{u_m=0})^{-1}+1}.\label{Eq:interface magnetism effect}
\end{equation}
This expression is of course consistent with Eqs.~(\ref{Eq:r matrix bilayer um}) and (\ref{Eq:r matrix FI with um}). The other scattering matrices are given by the constraints $1+\hat{r}_{\rm ex}=\sqrt{|k_z||\hat{K}_z|^{-1}}\hat{t}_{\rm ex}$, $1+\hat{r}_{\rm ex}'=\sqrt{|\hat{K}_z||k_z|^{-1}}\hat{t}_{\rm ex}'$, and $(1+\hat{r}_{\rm ex})k_z^{-1}=(1+\hat{r}_{\rm ex}')\hat{K}_z^{-1}$.

Equation~(\ref{Eq:interface magnetism effect}) allows for the exploration of interface magnetism effects up to any higher order in $u_m$ or $1/u_m$. By focusing on consequences of second term in Eq.~(\ref{Eq:interface magnetism effect}), one can look into the effects of proximity-induced magnetism on a given expression.

\section{Conclusion\label{Sec:conclusion}}

In summary, we develop a perturbation theory for scattering matrices to compute interfacial spin-orbit coupling effects in magnetic bilayers. We extend the two-dimensional Rashba model by embedding it in three-dimensional transport of electrons. We explicitly show that spin or charge current can be generated perpendicularly to an applied bias. Using this fact, we calculate current-induced (extrinsic) spin-orbit torque in terms of scattering amplitudes. For a given spin-orbit coupling Hamiltonian (like the Rashba form in our study), the resulting expressions from our theory are independent of details of the interface, so they are easily applicable for wide range of contexts. As demonstrations, we apply our formulas to various types of interfaces such as magnetic bilayers with bulk magnetism, those with interface magnetism, ferromagnetic insulators in contact with a nonmagnet, and topological insulators in contact with a ferromagnet.

For magnetic bilayers, we show that a dampinglike component can be on the same order of or larger than a fieldlike component, even without taking into account the Berry phase contribution and spin relaxation mechanisms. For the systems with insulating layers, we found that only a fieldlike component can arise, since a dampinglike component originates from a current across the interface. We also demonstrate that for finite bulk exchange coupling, the evanescent states that become important for the mismatched Fermi surfaces can give rise to the dominant contribution to spin-orbit torque.

Although we express the systems by analytic toy models, combining with first-principles calculations would enrich the implications of our theory significantly. We provide some remarks on possible generalization of our theory and future directions. Furthermore, we present other spin-orbit coupling phenomena, such as an in-plane current generation by a perpendicular bias (similar to the inverse spin Hall effect), a spin memory loss at the interface, and an anisotropic magnetoresistance (similar to the spin Hall magnetoresistance) in the appendices below. Our theory helps to characterize features of spin-orbit coupling phenomena for a given interface and further it provides insight on separating the roles of multiple sources of spin-orbit coupling effects such as spin Hall effect, interfacial spin-orbit coupling, and the magnetic proximity effect.

\textit{Note} During preparation of the manuscript, we found a recent report~\cite{Borge17arXiv} which uses a similar scattering formalism to our theory and describes several interface spin-orbit coupling phenomena, but focuses on a particular context, metallic bilayers without interface magnetism.

\begin{acknowledgments}
The authors acknowledge J. McClelland, P. Haney, and O. Gomonay, for critical reading of the manuscript. K.W.K acknowledges V.~Amin, and D.-S.~Han for fruitful discussion. K.W.K. was supported by the Cooperative Research Agreement between the University of Maryland and the National Institute of Standards and Technology, Center for Nanoscale Science and Technology (70NANB10H193), through the University of Maryland. K.W.K also acknowledges support by Basic Science Research Program through the National Research Foundation of Korea (NRF) funded by the Ministry of Education (2016R1A6A3A03008831). K.W.K and J.S. are supported by Alexander von Humboldt Foundation, the ERC Synergy Grant SC2 (No. 610115), and the Transregional Collaborative Research Center (SFB/TRR) 173 SPIN+X. K.J.L was supported by the National Research Foundation of Korea (2015M3D1A1070465, 2017R1A2B2006119). H.W.L. was supported by the SBS Foundation.
\end{acknowledgments}

\begin{appendix}
\section{Other physical consequences of interfacial spin-orbit coupling\label{Sec(A):Appendix A}}
\subsection{In-plane current induced by a perpendicular bias\label{Sec(A):results1A}}

The spin-orbit torque derived in the main text is essentially perpendicular spin current generation by in-plane charge current flow. Here we derive its Onsager counterpart. When a perpendicular bias (chemical potential difference) is applied, an in-plane current can be generated.

Suppose first that there is no spin-orbit coupling and note that the current operator is proportional to $\vec{k}$. If the system has rotational symmetry around $xy$ plane, all the scattering matrices must satisfy $\hat{r}_\vec{k}=\hat{r}_{-\vec{k}}$ for any in-plane $\vec{k}$ vector and similar relations for $\hat{r}'$, $\hat{t}$, and $\hat{t}'$. Thus, even if there is a perpendicular bias, any contribution from $\vec{k}$ to an in-plane current is canceled out by the opposite state $-\vec{k}$. Therefore, there is no in-plane current generation by a perpendicular bias.

However, the situation drastically changes when interfacial spin-orbit coupling is introduced. Here we present the perturbation result in the main text again.
\begin{subequations}
	\label{Eq(A):scattering matrices perturbation}
	\begin{align}
	\hat{t}_{\vec{k}}&=\hat{t}_{\vec{k}}^0+\frac{h_R}{2ik_z}\hat{t}_{\vec{k}, \rm ex}^0\hat{\vec{\sigma}}\cdot(\vec{k}\times\vhat{z})(\hat{1}_\vec{k}+\hat{r}_\vec{k}^0),\\
	\hat{t}_{\vec{k}}'&=\hat{t}_{\vec{k}}'^0+\frac{h_R}{2ik_z}(1+\hat{r}_{\vec{k}, \rm ex}^0)\hat{\vec{\sigma}}\cdot(\vec{k}\times\vhat{z})\hat{t}_\vec{k}'^0,\\
	\hat{r}_{\vec{k}}&=\hat{r}_{\vec{k}}^0+\frac{h_R}{2ik_z}(1+\hat{r}_{\vec{k}, \rm ex}^0)\hat{\vec{\sigma}}\cdot(\vec{k}\times\vhat{z})(\hat{1}_\vec{k}+\hat{r}_\vec{k}^0),\\
	\hat{r}_{\vec{k}}'&=\hat{r}_{\vec{k}}'^0+\frac{h_R}{2ik_z}\hat{t}_{\vec{k}, \rm ex}^0\hat{\vec{\sigma}}\cdot(\vec{k}\times\vhat{z})\hat{t}_\vec{k}'^0.
	\end{align}
\end{subequations}
The Rashba contributions are odd in $\vec{k}$. When they are multiplied by the current operator, the contributions from $\vec{k}$ and $-\vec{k}$ are no longer canceled out, thus an in-plane current can arise. Since the Rashba contribution is also proportional to the Pauli matrix vector, a charge bias will generate an in-plane spin current and a spin bias will generate an in-plane charge current. The latter has the same symmetry as the inverse spin Hall effect, implying that one needs to be careful when analyzing experiments~\cite{Mosendz10PRL,Mosendz10PRL,Weiler14PRL,Czeschka11PRL,Azevedo11PRB,Sanchez14PRL,Wang14PRL} using the inverse spin Hall effect as highlighted in \ocite{Wang16PRL}.

Here we derive explicit expressions of the current density at the interface. We do not assume $\vec{k}$ to be in-plane, thus our result will also recover the results of the magnetoelectric circuit theory. The current density at the normal metal along a unit vector $\vec{u}$ is calculated by
\begin{equation}
\tilde{j}_\vec{u}(z<0)=-\frac{e}{2}\Tr_\sigma[\rho\{v_u,\delta(\vec{r}_{\rm op}-\vec{r})\}\hat{\tilde{\sigma}}],
\end{equation}
where $v_u=(\hbar/ m_e i)\partial_u$ and $\rho$ is the density matrix, $\vec{r}_{\rm op}$ is the position operator, and $\vec{r}$ is the position $c$-number. $\tilde{\sigma}=(1, \vec{\sigma})$ is the four-dimensional Pauli matrix vector. $\tilde{j}_u$ is a four-dimensional vector whose zeroth component is the charge current along $\vec{u}$ and the other three components are the spin current along $\vec{u}$ with spin $x,y,z$ directions. Here and from now on, we denote any four-dimensional vector by the $\tilde{\cdot}$ notation.  As we develop in the main text, each of the eigenstates is written by a wave incident from the normal metal or a wave incident from the ferromagnet. Thus, they allow writing down the density matrix by a block-diagonal form. Using the notation of direct summation, $\rho=\rho^{\rm N}+\rho^{\rm F}$, where $\rho^{\rm N/F}$ are the density matrices block consisting of electrons incident from the normal metal/ferromegnet side. Thus we split the current into two terms: $\tilde{j}_\vec{u}=\tilde{j}_\vec{u}^{\rm N}+\tilde{j}_\vec{u}^{\rm F}$, where $\tilde{j}_\vec{u}^{\rm N/F}=-\frac{e}{2}\Tr_\sigma[\rho^{\rm N/F}\{v_u,\delta(\vec{r}_{\rm op}-\vec{r})\}\hat{\tilde{\sigma}}]$.

Let $\rho^{\rm N}=\sum_{\vec{k}\sigma'\sigma}f_{\vec{k},\sigma'\sigma}^{\rm N}|\vec{k}\sigma'\rangle\langle\vec{k}\sigma|$ where $f_{\vec{k},\sigma'\sigma}^{\rm N}$ is the $2\times2$ reduced density matrix. In a matrix form $\hat{f}_\vec{k}^{\rm N}$, each component is given by $f_{\vec{k},\sigma'\sigma}^{\rm N}=\xi_{\sigma'}^\dagger\hat{f}_\vec{k}^{\rm N}\xi_{\sigma}$. Since we consider a noncollinear spin injection from the normal metal, we allow for $\hat{f}_\vec{k}^{\rm N}$ having an off-diagonal component. By its definition, $\hat{1}_\vec{k}\hat{f}_\vec{k}^{\rm N}\hat{1}_\vec{k}=\hat{f}_\vec{k}^{\rm N}$, since there is no incident electrons out of the Hilbert space. The current at the normal metal from $\rho^{\rm N}$ is then calculated by the wave function Eq.~(\ref{Eq:psi^N(z<0)}). After some algebra,
\begin{equation}
\hat{j}_\vec{u}^{\rm N}(z<0)=-\frac{e\hbar}{m_eV}\sum_{\vec{k}}(\vec{k}\cdot\vec{u}\hat{f}_\vec{k}^{\rm N}+\vec{k}\cdot\bar{\vec{u}}\hat{r}_\vec{k}\hat{f}_\vec{k}^{\rm N}\hat{r}_\vec{k}^\dagger),\label{Eq(A):juN sum over k}
\end{equation}
where $\bar{\vec{u}}=(u_x, u_y, -u_z)$. $\hat{j}_\vec{u}^{\rm N/F}$ is a $2\times2$ matrix whose Pauli components are $(1/2)\tilde{j}_\vec{u}^{\rm N/F}$, that is, $\tilde{j}_\vec{u}^{\rm N/F}=\Tr_{\sigma}[\hat{\tilde{\sigma}}\hat{j}_\vec{u}^{\rm N/F}]$.

When a perpendicular bias is applied, the distribution function shifts. In the linear response regime, the distribution shift occurs only near the Fermi surface. To focus on the nonequilibrium current, we replace $\hat{f}_\vec{k}^{\rm N}=e\Delta \hat{\mu}^{\rm N}\hat{1}_\vec{k}\delta(E-E_F)$ where $\Delta\hat{\mu}^{\rm N}$ is the shift of the chemical potential of the normal metal due to the bias. To deal with the delta function easily, we convert the summation in Eq.~(\ref{Eq(A):juN sum over k}) to an integration: $\sum_\vec{k}\to(L/2\pi)\sum_{\vec{k}_\perp}\int dk_z$. By using $dE=(\hbar^2/m_e)k_zdk_z$, we convert the summation to an integration over energy. Due to the delta function, the energy integration is nothing but the integrand evaluated at the Fermi level. As a result, we obtain
\begin{equation}
\hat{j}_\vec{u}^{\rm N}(z<0)=-\frac{e^2L}{hV}\sum_{\vec{k}_\perp}\frac{1}{k_z}[\vec{k}\cdot\vec{u}\Delta\hat{\mu}^{\rm N}\hat{1}_\vec{k}+\vec{k}\cdot\bar{\vec{u}}\hat{r}_\vec{k}\Delta\hat{\mu}^{\rm N}\hat{r}_\vec{k}^\dagger]_{E=E_F},\label{Eq(A):juF z<0 (1)}
\end{equation}
where $L$ is the length along $z$ direction.

We use a similar method to obtain $\hat{j}_\vec{u}^{\rm F}$. There are three differences. First, we assume that there are no off-diagonal elements in $\hat{f}_\vec{k}^{\rm F}$ due to strong dephasing. Second, when we convert the summation by an integration, $\sum_\vec{k}\to(L/2\pi)\sum_{\vec{k}_\perp}\int dk_z^\sigma$, instead of $dk_z$, because the wave function is normalized by the incident wave. And then, we use $(\hbar^2/m_e)k_z^\sigma dk_z^\sigma=dE$. Third, the intervals of the integrations are different. The integral interval for $\hat{j}_\vec{u}^{\rm N}$ is $0<E<E_F$. However, in this case, the integral interval is $\sigma J<E<E_F$. However, since we focus on the Fermi surface contributions only, the lower bound of the energy does not matter. Omitting the algebra, we obtain
\begin{equation}
\hat{j}_\vec{u}^{\rm F}(z<0)=-\frac{e^2L}{hV}\Re\sum_{\vec{k}_\perp}\left[\frac{\vec{k}\cdot\bar{\vec{u}}}{|k_z|}e^{2\Im[k_z]z}\hat{t}_\vec{k}'\Delta\hat{\mu}^{\rm F}\hat{t}_\vec{k}'^\dagger\right]_{E=E_F}.\label{Eq(A):juF z<0 (2)}
\end{equation}
Now, the current right in the normal metal near the interface is given by the Pauli components of
\begin{equation}
\hat{j}_\vec{u}(z<0)=\hat{j}_\vec{u}^{\rm N}(z<0)+\hat{j}_\vec{u}^{\rm F}(z<0).\label{Eq(A):juF z<0}
\end{equation}

In a similar way, we obtain the expression of the current in the ferromagnet near the interface.
\begin{subequations}
\label{Eq(A):juF z>0}
\begin{equation}
\hat{j}_\vec{u}(z>0)=\hat{j}_\vec{u}^{\rm N}(z>0)+\hat{j}_\vec{u}^{\rm F}(z>0),
\end{equation}
where
\begin{widetext}
\begin{align}
\hat{j}_\vec{u}^{\rm N}(z>0)&=-\frac{e^2L}{hV}\Re\sum_{\vec{k}_\perp}\left[\hat{\vec{K}}\cdot\vec{u}\frac{e^{i\hat{K}_zz}}{\sqrt{|\hat{K}_z|}}\hat{t}_\vec{k}\Delta\hat{\mu}^{\rm N}t_\vec{k}^\dagger\frac{e^{-i\hat{K}_z^*z}}{\sqrt{|\hat{K}_z|}}\right]_{E=E_F},\\\hat{j}_\vec{u}^{\rm F}(z>0)&=-\frac{e^2L}{hV}\Re\sum_{\vec{k}_\perp}\hat{\vec{K}}\cdot\left\{\frac{e^{i\hat{K}_zz}}{\sqrt{|\hat{K}_z|}}\left[\bar{\vec{u}}\Delta\hat{\mu}^{\rm F}\hat{1}_{\vec{k}'}
+\vec{u}\hat{r}_\vec{k}'\Delta\hat{\mu}^{\rm F}\hat{r}_\vec{k}'^\dagger e^{-2\Im[\hat{K}_z]z}+(\bar{\vec{u}}e^{-2i\hat{K}_zz}\Delta\hat{\mu}^{\rm F}\hat{r}_\vec{k}'^\dagger+\vec{u}\hat{r}_\vec{k}'\Delta\hat{\mu}^{\rm F}e^{2i\hat{K}_zz})\right]\frac{e^{-i\hat{K}_zz}}{\sqrt{\hat{K}_z}}\right\}_{E=E_F},
\end{align}
\end{widetext}
\end{subequations}
where $\hat{\vec{K}}=(k_x, k_y, \hat{K}_z)$ is a vector consisting of $2\times2$ matrices. From Eqs.~(\ref{Eq(A):juF z<0 (1)})--(\ref{Eq(A):juF z>0}), one can compute the current near the interface for given (spin/charge) chemical potential excitation. As in the main text, we from now on omit the $[\cdots]_{E=E_F}$ and implicitly assume that the expressions are evaluated at the Fermi level.
	
We now simplify the expressions more. In Eq.~(\ref{Eq(A):juF z<0 (2)}), the $\Im[k_z]$ contribution originates from transmitted evanescent waves incident from the ferromagnet. For perpendicular transport, since $\vec{k}\cdot\bar{\vec{u}}/|k_z|$ is imaginary, there is no contribution from evanescent modes to a perpendicular current, consistently with the conservation of charge current. However, for in-plane transport, such a contribution can be nonzero. Note that the evanescent contribution dies after $1/k_z$ length scale. Since $1/k_F$ is shorter than the mean free path scale, the current is almost unmeasurable in experimental resolution. Thus, we neglect decaying contributions in Eqs.~(\ref{Eq(A):juF z<0 (2)}) and (\ref{Eq(A):juF z>0}). \footnote{The approximation is exact for perpendicular transport.}$^,$\footnote{If $k_\perp$ is sufficiently close to $k_F$, this does not hold. Therefore, this approximation is a crude approximation even for a low experimental resolution. However, we present this here since the approximation simplifies the expressions a lot.} We also neglect highly oscillating terms in Eq.~(\ref{Eq(A):juF z>0}). This is a common approximation to take into account dephasing of a transverse component to $\vec{m}$ in the ferromagnet. Then, we obtain
\begin{subequations}
\label{Eq(A):simplified current}
\begin{align}
\hat{j}_\vec{u}(z<0)&=-\frac{e^2L}{hV}\sum_{\vec{k}_\perp}\frac{\hat{1}_\vec{k}}{k_z}[\vec{k}\cdot\bar{\vec{u}}(\hat{r}_\vec{k}\Delta\hat{\mu}^{\rm N}\hat{r}_\vec{k}^\dagger+\hat{t}_\vec{k}'\Delta\hat{\mu}^{\rm F}\hat{t}_\vec{k}'^\dagger)\nonumber\\
&\phantom{=-\frac{e^2L}{hV}\sum_{\vec{k}_\perp}\frac{\hat{1}_\vec{k}}{k_z}}+\vec{k}\cdot\vec{u}\Delta\hat{\mu}^{\rm N}],\displaybreak[1]\\
\hat{j}_\vec{u}(z>0)&=-\frac{e^2L}{hV}\sum_{\vec{k}_\perp}\frac{\hat{1}_{\vec{k}'}}{\hat{K}_z}[\hat{\vec{K}}\cdot\vec{u}\Diag[\hat{r}_\vec{k}'\Delta\hat{\mu}^{\rm F}\hat{r}_\vec{k}'^\dagger+\hat{t}_\vec{k}\Delta\hat{\mu}^{\rm N}\hat{t}_\vec{k}^\dagger]]\nonumber\\
&\phantom{=-\frac{e^2L}{hV}\sum_{\vec{k}_\perp}\frac{\hat{1}_{\vec{k}'}}{\hat{K}_z}[}+\vec{\hat{K}}\cdot\bar{\vec{u}}\Delta\hat{\mu}^{\rm F},
\end{align}
\end{subequations}
where $\Diag[\cdots]=\sum_su_s[\cdots]u_s$ is the spin-diagonal part of a matrix. Physical meaning of this operation is the dephasing of a transverse component of spin in the ferromagnet.

We first take $\vec{u}=\vec{z}$ to see that Eq.~(\ref{Eq(A):simplified current}) is consistent with the conventional magnetoelectric circuit theory. It is easy to show that
\begin{equation}
\hat{j}_\vec{z}(z<0)=-\frac{e^2L}{hV}\sum_{k_\perp^2<k_F^2}[\Delta\hat{\mu}^{\rm N}-(\hat{r}_\vec{k}\Delta\hat{\mu}^{\rm N}\hat{r}_\vec{k}^\dagger+\hat{t}_\vec{k}'\Delta\hat{\mu}^{\rm F}\hat{t}_\vec{k}'^\dagger)],
\label{Eq(A):perpendicular transport0}
\end{equation}
which is exactly the result of the magnetoelectric circuit theory. One can also show with Eq.~(\ref{Eq(A):simplified current}) and the unitarity constraint developed in Sec.~\ref{Sec(A):Appendix Unitary} that $\hat{j}_\vec{z}(z>0)=\Diag[\hat{j}_\vec{z}(z<0)]$. This implies the continuity of electrical current across the interface, thus we do not need to keep $(z>0)$ or $(z<0)$. Following the procedure of the magnetoelectric circuit theory~\cite{Brataas00PRL,Brataas01EPJB}, we take $\Delta\hat{\mu}^{\rm N}=\Delta\mu_0^{\rm N}-\hat{\vec{\sigma}}\cdot\vec{s}\Delta\mu_s^{\rm N}$ where $\vec{s}$ is the direction of spin magnetic moment in the normal metal. $\vec{s}$ can be deviated from $\vec{m}$ when one considers noncollinear spin injection. We also take $\Delta\hat{\mu}^{\rm F}=\Delta\mu_0^{\rm F}-\hat{\vec{\sigma}}\cdot\vec{m}\Delta\mu_s^{\rm F}$.\footnote{In our model, the spin magnetic moment is antiparallel to the electron spin direction. Thus, we take $\Delta\mu_s^{\rm N/F}$ to have a negative sign to make a consistent notation with the previous theories.} The scattering matrices are taken by Eq.~(\ref{Eq(A):scattering matrices perturbation}). However, all the first order Rashba contributions are canceled out after summing over all transverse modes. They are odd in in-plane momentum $k_x$ or $k_y$, thus are canceled by an opposite contribution from $-k_x$ or $-k_y$. Therefore, we can discard the Rashba contributions and take only unperturbed scattering matrices, which are expanded by $\hat{r}_\vec{k}^{0}=r_\vec{k}^\da u_++r_\vec{k}^\ua u_-$ and so on. Then, spin and charge currents are given by
\begin{subequations}
\label{Eq(A):perpendicular transport}
\begin{align}
\frac{V}{L}\Tr_\sigma[\hat{j}_z]&=(G^{\ua\ua}+G^{\da\da})(\Delta\mu_0^{\rm F}-\Delta\mu_0^{\rm N})\nonumber\\
&\quad+(G^{\ua\ua}-G^{\da\da})(\Delta\mu_s^{\rm F}-\vec{m}\cdot\vec{s}\Delta\mu_s^{\rm N}),\displaybreak[1]\\
-\frac{V}{L}\Tr_\sigma[\hat{\vec{\sigma}}\hat{j}_z]&=(G^{\ua\ua}-G^{\da\da})(\Delta\mu_0^{\rm F}-\Delta\mu_0^{\rm N})\vec{m}\nonumber\\
&\quad+(G^{\ua\ua}+G^{\da\da})(\Delta\mu_s^{\rm F}-\vec{m}\cdot\vec{s}\Delta\mu_s^{\rm N})\vec{m}\nonumber\\
&\quad-2\Re\left[G^{\ua\da}[\vec{m}\times(\vec{s}\times\vec{m})+i\vec{s}\times\vec{m}]\right]\Delta\mu_s^{\rm N},
\end{align}
\end{subequations}
where $G^{ss'}=(e^2/h)\sum_{k_\perp^2<k_F^2}(1-r_\vec{k}^sr_\vec{k}^{s'*})$. $G^{\ua\ua/\da\da}$ is the interface conductance for spin majority/minority electrons and $G^{\ua\da}$ is the spin mixing conductance. Eq.~(\ref{Eq(A):perpendicular transport}) recovers all the results in the traditional theory.

We now take an in-plane $\vec{u}$. As we discuss above, non-Rashba contributions cannot generate an in-plane current. Using Eq.~(\ref{Eq(A):scattering matrices perturbation}) and collecting the first order contributions to $h_R$,
\begin{align}
\hat{j}_\vec{u}(z<0)&=-h_R\frac{e^2L}{hV}\Im\sum_{k_\perp^2<k_F^2}\frac{\vec{k}\cdot\vhat{u}}{k_z^2}(1+\hat{r}_\vec{k}^0)\hat{\vec{\sigma}}\cdot(\vec{k}\times\vhat{z})\nonumber\\
&\phantom{=-h_R\frac{e^2L}{hV}\Im\sum_{k_\perp^2<k_F^2}}
\times[(1+\hat{r}_\vec{k}^0)\Delta\hat{\mu}^{\rm N}\hat{r}_\vec{k}^{0\dagger}+\hat{t}_\vec{k}^0\Delta\hat{\mu}^{\rm F}\hat{t}_\vec{k}'^{0\dagger}].
\end{align}
When summing up over all transverse modes, one should consider all possible angles of $\vec{k}_\perp$ for a given magnitude. If the system has rotational symmetry, we can take an angle average of $(\vec{k}\cdot\vec{u})[\hat{\vec{\sigma}}\cdot(\vec{k}\times\vec{z})]$ by integrating over the in-plane angle from $-\pi$ to $\pi$ and dividing the result by $2\pi$. Then, we obtain $(k_\perp^2/2)\hat{\vec{\sigma}}\cdot(\vec{u}\times\vec{z})$. After the angle average,
\begin{align}
\hat{j}_\vec{u}(z<0)&=-h_R\frac{e^2L}{2hV}\Im\sum_{k_\perp^2<k_F^2}\frac{E_\perp}{E_F-E_\perp}(1+\hat{r}_\vec{k}^0)\hat{\vec{\sigma}}\cdot(\vec{u}\times\vhat{z})\nonumber\\
&\phantom{=-h_R\frac{e^2L}{hV}\Im\sum_{k_\perp^2<k_F^2}}
\times[(1+\hat{r}_\vec{k}^0)\Delta\hat{\mu}^{\rm N}\hat{r}_\vec{k}^{0\dagger}+\hat{t}_\vec{k}^0\Delta\hat{\mu}^{\rm F}\hat{t}_\vec{k}'^{0\dagger}].
\end{align}
Here $E_\perp=\hbar^2k_\perp^2/2m_e$ and we used $k_\perp^2/k_z^2=E_\perp/(E_F-E_\perp)$. Then, the in-plane spin and charge currents are given by its Pauli components.
\begin{widetext}
\begin{subequations}
\begin{align}
\frac{V}{L}\Tr_\sigma[\hat{j}_u(z<0)]
&=-\frac{(\vhat{u}\times\vhat{z})}{2i}\cdot\left[(G_{Rt}^{\ua\ua\ua}-G_{Rt}^{\da\da\da})(\Delta\mu_0^{\rm F}-\Delta\mu_0^{\rm N})\vec{m}+(G_{Rt}^{\ua\ua\ua}+G_{Rt}^{\da\da\da})(\Delta\mu_s^{\rm F}-\vec{m}\cdot\vec{s}\Delta\mu_s^{\rm N})\vec{m}\right.\nonumber\\
&\phantom{=-\frac{(\vhat{u}\times\vhat{z})}{2i}\cdot[}\left.+(G_{Rr}^{\ua\da\da}-G_{Rr}^{\da\ua\ua*})\Delta\mu_s^{\rm N}[\vec{m}\times(\vec{s}\times\vec{m})+i\vec{s}\times\vec{m}]\right]+\mathrm{c.c.},\label{Eq(A):charge current generation}\displaybreak[2]\\
-\frac{V}{L}\Tr_\sigma[\hat{\vec{\sigma}}\hat{j}_u(z<0)]&=-\vec{m}\frac{(\vhat{u}\times\vhat{z})}{2i}\cdot\left[(G_{Rt}^{\ua\ua\ua}+G_{Rt}^{\da\da\da})(\Delta\mu_0^{\rm F}-\Delta\mu_0^{\rm N})\vec{m}+(G_{Rt}^{\ua\ua\ua}-G_{Rt}^{\da\da\da})(\Delta\mu_s^{\rm F}-\vec{m}\cdot\vec{s}\Delta\mu_s^{\rm N})\vec{m}\right.\nonumber\\
&\phantom{=\vec{m}\frac{(\vhat{u}\times\vhat{z})}{2i}\cdot[}\left.~-(G_{Rr}^{\ua\da\da}+G_{Rr}^{\da\ua\ua*})\Delta\mu_s^{\rm N}[\vec{m}\times(\vec{s}\times\vec{m})+i\vec{s}\times\vec{m}]\right]\nonumber\\
&\quad-\frac{1}{2i}\left[(G_{Rt}^{\ua\da\da}-G_{Rt}^{\da\ua\ua*})\Delta\mu_0^{\rm F}-(G_{Rt}^{\ua\da\da}+G_{Rt}^{\da\ua\ua*})\Delta\mu_s^{\rm F}+(G_{Rr}^{\ua\da\da}-G_{Rr}^{\da\ua\ua*})\Delta\mu_0^{\rm N}-\vec{m}\cdot\vec{s}(G_{Rr}^{\ua\da\da}+G_{Rr}^{\da\ua\ua*})\Delta\mu_s^{\rm N}\right]\nonumber\\
&\phantom{\quad-}\times\{\vec{m}\times[(\vhat{u}\times\vhat{z})\times\vec{m}]-i\vec{m}\times(\vhat{u}\times\vhat{z})\}-\frac{(\vhat{u}\times\vhat{z})\cdot\vec{m}}{2i}(G_{Rr}^{\ua\ua\da}+G_{Rr}^{\da\da\ua*})\Delta\mu_s^{\rm N}[\vec{m}\times(\vec{s}\times\vec{m})+i\vec{s}\times\vec{m}]+\mathrm{c.c}.
\end{align}
\end{subequations}
\end{widetext}
where c.c. refers to complex conjugate of all terms in front of it. In the conductances $G_{Rr}^{ss's''}$ and $G_{Rt}^{ss's''}$, the subscript $R$ refers to Rashba contributions, and $r$ and $t$ refer to contributions from reflection and transmission. The explicit expressions in terms of unperturbed scattering matrices are
\begin{subequations}
\begin{align}
G_{Rr}^{ss's''}&=-h_R\frac{e^2}{2h}\sum_{\vec{k}_\perp}\frac{E_\perp}{E_F-E_\perp}(1+r_\vec{k}^s)(1+r_\vec{k}^{s'})r_\vec{k}^{s''*},\\
G_{Rt}^{ss's''}&=-h_R\frac{e^2}{2h}\sum_{\vec{k}_\perp}\frac{E_\perp}{E_F-E_\perp}(1+r_\vec{k}^s)t_\vec{k}'^{s'}t_\vec{k}'^{s''*}.
\end{align}
\end{subequations}

Equation~(\ref{Eq(A):charge current generation}) clearly demonstrates a charge current is generated by a spin chemical potential bias. For the case of collinear transport ($\vec{s}=\vec{m}$), the charge current along $\vec{u}=\vec{x}$ is proportional to $m_y$. Since the inverse spin Hall effect is also dependent on $m_y$, it requires a careful analysis. Simultaneous description of the inverse spin Hall effect and interfacial spin-orbit coupling would be a future challenge.

As passing remarks, we present some properties of $G_{Rr/t}^{ss's''}$. First, there are three indices. This is because there is interfacial spin-orbit coupling as an additional spin scattering source. A complete description requires three indices; first one for the incident spin, second one for the scattering due to interfacial spin-orbit coupling, and third one for the transmitted spin. Second, $G_{Rr/t}^{ss's''}$ is symmetric under exchange between $s$ and $s'$. Third, the unitarity constraint in Sec.~\ref{Sec(A):Appendix Unitary} implies $\Im[G_{Rr}^{sss}+\Im[G_{Rt}^{sss}]=0$. This is shown by using $|r_\vec{k}^s|^2+|t_\vec{k}'^s|^2=1$ to derive $G_{Rr}^{sss}+G_{Rt}^{sss}=-(h_Re^2/2h) \sum_{k_\perp^2<k_F^2}[(E_\perp)/(E_F-E_\perp)]|1+r_\vec{k}^s|^2$. This constraint guarantees the absence of a charge current generation at equilibrium.

We also remark that in-plane current can be discontinuous at the interface in the presence of interfacial spin-orbit coupling. To see this, we define a discontinuity matrix by $\Delta\hat{j}_\vec{u}=\hat{j}_\vec{u}(z>0)-\Diag[\hat{j}_\vec{u}(z<0)]$. This is zero for $\vec{u}=\vec{z}$, not for an in-plane $\vec{u}$. After some algebra, we obtain
\begin{widetext}
\begin{subequations}
\begin{align}
\frac{V}{L}\Tr_\sigma[\Delta\hat{j}_\vec{u}]&=\frac{(\vhat{u}\times\vhat{z})}{2i}\cdot\left[(\Delta G_h^{\ua\ua}-G_{R\Delta}^{\da\da})(\Delta\mu_0^{\rm F}-\Delta\mu_0^{\rm N})\vec{m}+(G_{R\Delta}^{\ua\ua}+G_{R\Delta}^{\da\da})(\Delta\mu_s^{\rm F}-\vec{m}\cdot\vec{s}\Delta\mu_s^{\rm N})\vec{m}-2G_{R\Delta}^{\ua\da}\vec{m}\times(\vec{s}\times\vec{m})\Delta\mu_s^{\rm N}\right]+\mathrm{c.c.},\\
-\frac{V}{L}\Tr_\sigma[\hat{\vec{\sigma}}\Delta\hat{j}_\vec{u}]&=\frac{(\vhat{u}\times\vhat{z})}{2i}\cdot\left[(G_{R\Delta}^{\ua\ua}+G_{R\Delta}^{\da\da})(\Delta\mu_0^{\rm F}-\Delta\mu_0^{\rm N})\vec{m}+(G_{R\Delta}^{\ua\ua}-G_{R\Delta}^{\da\da})(\Delta\mu_s^{\rm F}-\vec{m}\cdot\vec{s}\Delta\mu_s^{\rm N})\vec{m}+2iG_{R\Delta}^{\ua\da}\vec{s}\times\vec{m}\Delta\mu_s^{\rm N}\right]+\mathrm{c.c.},
\end{align}
\end{subequations}
\end{widetext}
where
\begin{equation}
G_{R\Delta}^{ss'}=-h_R\frac{e^2}{2h}\sum_{k_\perp^2<k_F^2}\frac{E_\perp}{E_F-E_\perp}(1+r_\vec{k}^s)(1+r_\vec{k}^{s'}).\label{Eq(A):discontinuity conductance}
\end{equation}

Before closing the section, we mention that the currents are proportional to $L$, the size of the system. In reality, the contributions will relax on the length scale of the mean free path $\lambda$. Thus, if $L\gg \lambda$, $L$ in the above expressions should be replaced by $\lambda$ when one takes into account the bulk scattering.

In the main text and this section, we have demonstrated that a bias can generate a current perpendicular to the applied bias direction, not affecting its longitudinal transport. These results are first order perturbation theory. In the following two sections, we calculate second order effects in $h_R$ to examine the effects of interfacial spin-orbit coupling on \emph{longitudinal transport}. Each section deals with a perpendicular bias to the interface and an in-plane bias respectively.

\subsection{Spin memory loss and spin torque from collinear spin injection\label{Sec(A):results1C}}

Equation~(\ref{Eq(A):perpendicular transport}) describes the generation of a perpendicular current to the interface in the presence of a perpendicular bias. Spin flip at the interface due to interfacial spin-orbit coupling vanishes due to rotation symmetry around the $xy$ plane. However, if we consider second order contributions, the result can change. Below we demonstrate, a spin-up (down) current can be generated by a spin-down (up) bias. This means that interfacial spin-orbit coupling flips the spin at the interface even when we consider a collinear transport. We interpret this as spin memory loss at the interface.

We start from Eq.~(\ref{Eq(A):perpendicular transport0}). The second order expansion is given by Eq.~(\ref{Eq:multi generalization of r}). For simplicity, we consider only collinear transport with perpendicular magnetization ($\vec{s}=\vec{m}=\vec{z}$). For $\vec{s}\ne\vec{m}$, the coefficient of spin torque in Eq.~(\ref{Eq(A):perpendicular transport}) will also change. For $\vec{m}\ne\vec{z}$, the result will depend on the direction of the magnetization, leading to a current-perpendicular magnetoresistance. Below we make more remarks on this case.

Writing down the expression explicitly, one realizes that the Rashba contributions arises in the form of $\hat{\kappa}_R \hat{A} \hat{\kappa}_R$. For instance, there is a contribution proportional to $4k_z^2\hat{G}[\hat{\kappa}_R\hat{G}\Delta\hat{\mu}^{\rm N}\hat{G}^\dagger\hat{\kappa}_R]\hat{G}^\dagger$. The angle average over all transverse modes simplifies such expressions a lot. From Eq.~(\ref{Eq:kappa_R expression}) and some algebra, angle average of $\hat{\kappa}_R \hat{A} \hat{\kappa}_R$ for an arbitrary diagonal matrix $\hat{A}$ is also diagonal:
\begin{equation}
\hat{\kappa}_R\left(
\begin{array}{cc}
a_1&0\\
0&a_2
\end{array}
\right)\hat{\kappa}_R\to h_R^2k_\perp^2\left(
\begin{array}{cc}
a_2&0\\
0&a_1
\end{array}
\right).\label{Eq(A):kappa second order angle average}
\end{equation}
We emphasize that the diagonal components are exchanged. As we show below, this exchange results in mixing of spin-up and spin-down components. As a result, a spin-up (down) bias can generate spin-down (up) current.

To see spin flip at the interface clearly, we use up/down ($\ua$/$\da$) notations rather than the spin/charge ($0$/$s$) notations used in Sec.~\ref{Sec(A):results1A}. We expand $\Delta\hat{\mu}^{\rm N}=\Delta\mu_\da^{\rm N}u_++\Delta\mu_\ua^{\rm N}u_-$ and similarly for $\Delta\hat{\mu}^{\rm F}$.\footnote{The relation between $\Delta\mu_{\ua\da}^{\rm N}$ and $\Delta\mu_{0/s}^{\rm N}$ is $\Delta\mu_{\ua\da}^{\rm N}=\Delta\mu_0^{\rm N}\pm\Delta\mu_s^{\rm N}$ and a similar relation holds for $\Delta\hat{\mu}^{\rm F}$ and $I_{\ua\da}$.} We also define $I_{\ua\da}=(V/L)\Tr_\sigma[u_\mp\hat{j}_z]$, which are spin-up/down currents. We simply denote $\Delta \mu_{\ua/\da}$, $\Delta\mu_{\ua\da}^{\rm F}-\Delta\mu_{\ua\da}^{\rm N}$, the spin chemical potential difference across the interface. After some algebra,
\begin{subequations}
\begin{align}
\left(\begin{array}{c}
I_\ua\\
I_\da
\end{array}\right)&=\left(
\begin{array}{cc}
G^{\ua\ua}-2h_R\Re G_{Rr}^{\da\ua\ua}& G_{\rm flip} \\
G_{\rm flip} & G^{\da\da}-2h_R\Re G_{Rr}^{\ua\da\da}
\end{array} \right)\left(\begin{array}{c}
\Delta\mu_\ua\\
\Delta\mu_\da
\end{array}\right),\\
G_{\rm flip}&=h_R^2\frac{e^2}{2h}\sum_{k_\perp^2<k_F^2}\frac{E_\perp}{E_F-E_\perp}|1+r_\vec{k}^{\ua}|^2|1+r_\vec{k}^\da|^2.
\end{align}
\end{subequations}
Here $G^{ss}$ are the unperturbed interface conductances derived from the magnetoelectric circuit theory. The corrections to the diagonal terms are the second order corrections\footnote{Since $G_{Rr}^{ss's''}$ is already in first order in $h_R$, the corrections are in second order.} from interfacial spin-orbit coupling. These diagonal terms simply describes a spin-up/down current generation by a spin-up/down bias. On the other hand, $G_{\rm flip}$ describes spin-flipping contributions: A spin-up (down) bias generates spin down (up) current.

If $\vec{m}\ne\vec{z}$, Eq.~(\ref{Eq(A):kappa second order angle average}) does not hold. Even the result is not diagonal.\footnote{In Sec.~\ref{Sec(A):results1E}, we demonstrate how an angle average can have off-diagonal components for a general $\vec{m}$.} The existence of an off-diagonal element implies that one cannot use the two-current model and there arises a spin current whose spin direction is transverse to the magnetization. This means that a spin torque (depending on $m_z$) can be generated even for a collinear spin injection. Due to complexity of the expressions, we do now show it here, but the expression is given by exactly the same procedure [except Eq.~(\ref{Eq(A):kappa second order angle average})]. Although this is a second order effect, it would be valuable if this spin torque is experimentally realized.

The spin memory loss and spin torque we illustrate above do not violate the conservation of angular momentum. The source of the angular momentum is nothing but the lattice at the interface. Spin-orbit coupling at the interface pumps orbital angular momentum from (to) the lattice to (from) the spin-magnetization system.

\subsection{Anisotropic magnetoresistance\label{Sec(A):results1E}}
In the previous section, we examine second order effects for perpendicular transport to the interface. In this section, we consider in-plane transport. We below show that the electrical resistance depends on the direction of magnetization. When a charge current is flowing along $\vec{x}$, the resistance includes terms proportional to $m_x^2$ and $m_y^2$. The former has the same symmetry with the conventional anisotropic magnetoresistance from ferromagnetic bulk, but can have the opposite sign. The latter has the same symmetry as the spin Hall magnetoresistance~\cite{Huang12PRL,Nakayama13PRL,Hahn13PRB,Chen13PRB,Althammer13PRB,Kim16PRL}.

For simplicity we consider a current mainly flowing in the normal metal. When a charge current is flowing in the normal metal, the distribution function is shifted: $f_\vec{k}^{\rm N}=f_\vec{k}^{0,\rm N}+(eE\hbar/m_e)k_x\delta(E-E_F)$ where $\vec{x}$ is the direction of the current flow. Or equivalently, $\Delta\hat{\mu}^{\rm N}=(E\hbar \tau/m_e)k_x$. Putting this and $\vec{u}=\vec{x}$ into Eq.~(\ref{Eq(A):juF z<0 (1)}) and taking trace over the spin space gives the charge current due to the distribution shift.
\begin{equation}
\Tr_{\sigma}[\hat{j}_x^{\rm N}]=-\frac{e^2E\tau L}{2\pi m_eV}\Tr_{\sigma}\sum_{k_\perp^2<k_F^2}\frac{k_x^2}{k_z}(1+\hat{r}_\vec{k}\hat{r}_\vec{k}^\dagger).\label{Eq(A):charge current in-plane}
\end{equation}
The first term in the parenthesis ($1$) is the conventional electrical current and the second term ($\hat{r}_\vec{k}\hat{r}_\vec{k}^\dagger$) is the Rashba contribution. The second order expansion is given by Eq.~(\ref{Eq:multi generalization of r}). First order Rashba contributions are odd in $\vec{k}_\perp$ so are all canceled out after summing up over all transverse modes.

To compute second order contributions, by the same reason in Sec.~\ref{Sec(A):results1C}, we take angle average of the expression $\hat{\kappa}_R\hat{A}\hat{\kappa}_R$ for a diagonal $\hat{A}$.
\begin{align}
k_x^2\hat{\kappa}_R\left(
\begin{array}{cc}
a_1&0\\
0&a_2
\end{array}
\right)\hat{\kappa}_R&\to h_R^2\frac{k_\perp^4}{2}\left(
\begin{array}{cc}
a_2&0\\
0&a_1
\end{array}
\right)\nonumber\\&\quad+(a_2-a_1)\frac{h_R^2k_\perp^4m_z}{8m_\parallel}(m_x^2+3m_y^2)\hat{\sigma}_x\nonumber\\
&\quad+(a_2-a_1)\frac{h_R^2k_\perp^4}{4m_\parallel}m_xm_y\hat{\sigma}_y\nonumber\\&\quad+(a_2-a_1)\frac{h_R^2k_\perp^4}{8}(m_x^2+3m_y^2)\hat{\sigma}_z
,\label{Eq(A):angle average MR}
\end{align}
after angle average. Here $m_\parallel=\sqrt{m_x^2+m_y^2}$.

In charge transport, the result is given by taking the trace. Thus the diagonal terms (the first and last terms) only matter. Due to the existence of $(m_x^2+3m_y^2)$ contribution in front of $\hat{\sigma}_z$, a magnetoresistance proportional to $(m_x^2+3m_y^2)$ can arise. If the reflection matrix is spin independent, the magnetoresistance is zero since $\Tr[\sigma_z]=0$. However, if the reflection matrix is spin dependent, the contribution can survive. In a previous work~\cite{Zhang15PRB}, a magnetoresistance proportional to $(m_x^2+3m_y^2)$ is reported, but is proportional to $\tau^\ua-\tau^\da$, thus a current should flow in the ferromagnet. This is because their model assumes an equal-Fermi-surface model. On the other hand, we take into account more generalized situation and demonstrate that a magnetoresistance $\propto (m_x^2+3m_y^2)$ can arise if $r_\vec{k}^\ua\ne r_\vec{k}^\da$.

Another remark is that the term $3m_y^2$ has the same symmetry as the spin Hall magnetoresistance. Therefore, the spin Hall magnetoresistance should also be carefully analyzed since there is an interface contribution. There are several theoretical and experimental reports~\cite{Grigoryan14PRB,Zhang15PRB,Bijl12PRB,Nakayama16PRL} that interface Rashba effect can give rise to a magnetoresistance having the same symmetry as the spin Hall magnetoresistance. Another difference is that the Rashba contribution has a $m_x^2$ contribution as well. The contribution has the same symmetry as the conventional anisotropic magnetoresistance, but below we show that it can have the opposite sign (we call it a \emph{negative} magnetoresistance). Therefore, observing the spin-Hall-like magnetoresistance and a negative anisotropic magnetoresistance of a similar orders of magnitude will shed light on separating the bulk effects and the interfacial effects.

To complete our analysis, we explicitly calculate the current [Eq.~(\ref{Eq(A):charge current in-plane})] with Eqs.~(\ref{Eq:multi generalization of r}) and (\ref{Eq(A):angle average MR}). After some algebra,
\begin{subequations}
\label{Eq(A):MR}
\begin{align}
\Tr_{\sigma}[\hat{j}_x^{\rm N}]&=j_{\rm conv}+j_{\rm nonMR}+j_{\rm MR}(m_x^2+3m_y^2),\\
j_{\rm conv}&=-\frac{e^2E\tau L}{4\pi m_eV}\sum_{k_\perp^2<k_F^2,s=\ua,\da}\frac{k_\perp^2}{k_z}(1+|r_\vec{k}^s|^2),\\
j_{\rm nonMR}&=-h_R^2\frac{e^2E\tau L}{8\pi m_eV}\nonumber\\&\quad\times\Re\sum_{k_\perp^2<k_F^2}\frac{k_\perp^4}{k_z^3}\rho_\vec{k}^\ua \rho_\vec{k}^\da(\rho_\vec{k}^{\ua*}\rho_\vec{k}^{\da*}+\rho_\vec{k}^\da r_\vec{k}^{\da*}+\rho_\vec{k}^\ua r_\vec{k}^{\ua*}),\\
j_{\rm MR}&=-h_R^2\frac{e^2E\tau L}{32\pi m_eV}\Re\sum_{k_\perp^2<k_F^2}[(|\rho_\vec{k}^\ua|^2-|\rho_\vec{k}^\da|^2)^2\nonumber\\&\quad-2(\rho_\vec{k}^\ua-\rho_\vec{k}^\da)(\rho_\vec{k}^{\ua2}r_\vec{k}^{\ua*}-\rho_\vec{k}^{\da2}r_\vec{k}^{\da*})],
\end{align}
\end{subequations}
where $\rho_\vec{k}^{\ua\da}=1+r_\vec{k}^{\ua\da}$. Here $j_{\rm conv}$ is the non-Rashba contribution, $j_{\rm nonMR}$ is the second order Rashba correction that is independent of $\vec{m}$, and $j_{\rm MR}$ is the magnetoresistance. If the summand in $j_{\rm MR}$ is positive, the magnitude of the current increases when $m_x^2$ increases, which is the negative magnetoresistance. Below we discuss when a negative magnetoresistance arises.

For magnetic bilayers, if bulk magnetism is dominant, the reflection matrix is real [See Eq.~(\ref{Eq:reflection bulk magnetism_r})]. The reality of $r_\vec{k}^{\ua\da}$ allows simplifying the summand in $j_{\rm MR}$ as
\begin{align}
&(|\rho_\vec{k}^\ua|^2-|\rho_\vec{k}^\da|^2)^2-2\Re[(\rho_\vec{k}^\ua-\rho_\vec{k}^\da)(\rho_\vec{k}^{\ua2}r_\vec{k}^{\ua*}-\rho_\vec{k}^{\da2}r_\vec{k}^{\da*})])\nonumber\\&
=(r_\vec{k}^\ua-r_\vec{k}^\da)^2(2-r_\vec{k}^{\ua2}-r_\vec{k}^{\da2})\ge 0.
\end{align}
For ferromagnetic insulators, one can deduce $j_{\rm MR}=0$. This is consistent with the Landauer-B\"{u}ttiker formalism. Since the ferromagnet is insulating, the conductance (thus resistance) is determined by the number of transverse modes in the normal metal, and is independent of the magnetization direction. This is a general result which holds unless the translational symmetry along the current flowing direction is broken or there is a spin-dependent scattering source (for example, the spin Hall effect generating spin Hall magnetoresistance). The case for the topological insulators cannot be described by Eq.~(\ref{Eq(A):MR}) since we assume that a current is flowing in the normal metal. But we mention that it was studied in a previous report~\cite{Zhang15PRB}. For complex $r_\vec{k}^{\ua\da}$ in metallic magnetic bilayers, we found no a priori argument which guarantees the sign of $j_{\rm MR}$.



\section{Unitarity constraint of the scattering amplitudes\label{Sec(A):Appendix Unitary}}
Provided that all waves are propagating, charge conservation implies the
following unitarity constraint: For the following scattering matrix
\begin{equation}
S_\vec{k}=\left(
    \begin{array}{cc}
      \hat{r}_\vec{k} & \hat{t}_\vec{k}' \\
      \hat{t}_\vec{k} & \hat{r}_\vec{k}' \\
    \end{array}
  \right),
\end{equation}
$S_\vec{k}^\dagger S_\vec{k}=1$ holds. Up to this point, the expressions are conventional. However, this works only if all waves are propagating, thus it must be generalized to the extended space. In this section, we omit the subscript `ex' for simplicity.

Let the $\sigma=+1$ band be evanescent in the ferromagnet. As far as unitarity is concerned, some of matrices should be projected into $\sigma=-1$. Note that $S_\vec{k}$ connects incoming waves to outgoing waves.
\begin{equation}
\left(
  \begin{array}{c}
    \psi_{\vec{k}}^{N,\rm out} \\
    \psi_{\vec{k}}^{F,\rm out} \\
  \end{array}\right)=S_\vec{k}\left(
    \begin{array}{c}
    \psi_{\vec{k}}^{N,\rm in} \\
    \psi_{\vec{k}}^{F,\rm in} \\
  \end{array}
\right)
\end{equation}
The transmitted outgoing waves for $\sigma=+1$ electrons are dropped. Thus
$\hat{t}_\vec{k}\to u_-\hat{t}_\vec{k}$ and $\hat{r}_\vec{k}'\to
u_-\hat{r}_\vec{k}'$. In addition, there is no contribution incident from FM
for $\sigma=+1$ electrons. Thus $\hat{t}_\vec{k}'\to \hat{t}_\vec{k}'u_-$ and $u_-\hat{r}_\vec{k}'\to u_-\hat{r}_\vec{k}'u_-$. Thus the projected scattering matrix is
\begin{equation}
S_\vec{k}^{\rm proj}=\left(
    \begin{array}{cc}
      \hat{r}_\vec{k} & \hat{t}_\vec{k}'u_- \\
      u_-\hat{t}_\vec{k} & u_-\hat{r}_\vec{k}'u_- \\
    \end{array}
  \right).
\end{equation}
Lastly, the unitarity constraint is given by the conservation of electrical charge. Neglecting $\sigma=+1$ in the ferromagnet, the unitarity constraint is given by
\begin{equation}
(S_\vec{k}^{\rm proj})^\dagger S_\vec{k}^{\rm proj}=\left(
                                                  \begin{array}{cc}
                                                    1 & 0 \\
                                                    0 & u_- \\
                                                  \end{array}
                                                \right).
\end{equation}

Let only $\sigma=-1$ band be propagating. There is no
contribution from $\hat{r}_\vec{k}$, $\hat{t}_\vec{k}$, and $\hat{t}_\vec{k}'$ for the unitarity constraint.
\begin{equation}
S_\vec{k}^{\rm proj}=\left(
    \begin{array}{cc}
      0 & 0 \\
      0 & u_-\hat{r}_\vec{k}'u_- \\
    \end{array}
  \right),
\end{equation}
and the unitarity constraint is given by
\begin{equation}
(S_\vec{k}^{\rm proj})^\dagger S_\vec{k}^{\rm proj}=\left(
                                                  \begin{array}{cc}
                                                    0 & 0 \\
                                                    0 & u_- \\
                                                  \end{array}
                                                \right).
\end{equation}
The above relations also hold when the order of $(S_\vec{k}^{\rm
proj})^\dagger$ and $S_\vec{k}^{\rm proj}$ is reversed.

The three expressions can be combined by means of the projection
matrices.
\begin{gather}
S_\vec{k}^{\rm proj}=\left(
    \begin{array}{cc}
      \hat{1}_\vec{k}\hat{r}_\vec{k} & \hat{1}_\vec{k}\hat{t}_\vec{k}' \\
      \hat{1}_\vec{k}'\hat{t}_\vec{k} & \hat{1}_\vec{k}'\hat{r}_\vec{k}' \\
    \end{array}
  \right),\\
(S_\vec{k}^{\rm proj})^\dagger S_\vec{k}^{\rm proj}=S_\vec{k}^{\rm proj}(S_\vec{k}^{\rm proj})^\dagger=\left(
\begin{array}{cc}
\hat{1}_\vec{k} & 0 \\
0 & \hat{1}_\vec{k}' \\
\end{array}
\right).
\end{gather}
Explicitly calculating each component gives the unitarity constraint of scattering matrices in our theory.
\begin{subequations}
\begin{align}
\hat{r}_\vec{k}^\dagger\hat{1}_\vec{k}\hat{r}_\vec{k}+\hat{t}_\vec{k}^\dagger\hat{1}_\vec{k}'\hat{t}_\vec{k}&=\hat{1}_\vec{k},\\
\hat{r}_\vec{k}'^\dagger\hat{1}_\vec{k}'\hat{r}_\vec{k}'+\hat{t}_\vec{k}'^\dagger\hat{1}_\vec{k}\hat{t}_\vec{k}'&=\hat{1}_\vec{k}',\\
\hat{r}_\vec{k}^\dagger\hat{1}_\vec{k}\hat{t}_\vec{k}'+\hat{t}_\vec{k}^\dagger\hat{1}_\vec{k}'\hat{r}_\vec{k}'&=0,
\end{align}
\end{subequations}
and
\begin{subequations}
\begin{align}
\hat{1}_\vec{k}(\hat{r}_\vec{k}\hat{r}_\vec{k}^\dagger+\hat{t}_\vec{k}'\hat{t}_\vec{k}'^\dagger)\hat{1}_\vec{k}&=\hat{1}_\vec{k},\\
\hat{1}_\vec{k}'(\hat{r}_\vec{k}'\hat{r}_\vec{k}'^\dagger+\hat{t}_\vec{k}\hat{t}_\vec{k}^\dagger)\hat{1}_\vec{k}'&=\hat{1}_\vec{k}',\\
\hat{1}_\vec{k}(\hat{r}_\vec{k}\hat{t}_\vec{k}^\dagger+\hat{t}_\vec{k}'\hat{r}_\vec{k}'^\dagger)\hat{1}_\vec{k}'&=0.
\end{align}
\end{subequations}

\section{Absence of a bound state\label{Sec(A):Appendix B}}

We show that the delta-function spin-orbit coupling potential at $z=0$ does not create a bound state unless its magnitude is beyond a perturbative regime. For mathematical simplicity, we take a simpler model in which electrons are subject to the largest magnitude of the delta function. A bound state is most likely to exist in this situation. The maximum magnitude of the delta function is $(\hbar^2/2m_e)h_R k_\perp^{\rm max}$ where $k_\perp^{\rm max}$ is the maximum value of $\sqrt{k_x^2+k_y^2}$. Then, the Hamiltonian becomes
\begin{equation}
H=-\frac{\hbar^2\nabla^2}{2m_e}\pm J\Theta(z)-\frac{\hbar^2}{2m_e}h_Rk_\perp^{\rm max}\delta(z),
\end{equation}
where $\pm$ refers to each spin band. Let the bound state wave function be $e^{q_1z}$ for $z<0$ and $e^{-q_2z}$ for $z>0$. Then both $q_1$ and $q_2$ should be positive. The relation between $q_1$ and $q_2$ is $-\hbar^2q_1^2/2m_e=\pm J-\hbar^2q_2^2/2m_e$, or equivalently
\begin{equation}
(q_2-q_1)(q_2+q_1)=\pm \frac{2m_eJ}{\hbar^2}.
\end{equation}
Now, the derivative mismatching condition from the delta-function potential is
\begin{equation}
\frac{\hbar^2}{2m_e}(q_2+q_1)=\frac{\hbar^2}{2m_e}h_Rk_\perp^{\rm max}.
\end{equation}
Combining the two conditions,
\begin{equation}
2q_{1/2}=h_Rk_\perp^{\rm max}\pm\frac{Jm_e}{\hbar^2k_\perp^{\rm max}h_R},
\end{equation}
where the choice of the sign $\pm$ that corresponds to 1 or 2 is ambiguous but it is not necessary to be determined.

A necessary condition for $q_1$ and $q_2$ being positive is that $q_1q_2$ is positive. Thus, we obtain
\begin{equation}
|h_Rk_\perp^{\rm max}|>\sqrt{\frac{2m_eJ}{\hbar^2}}.\label{Eq(A):Bound state condition}
\end{equation}
For a perturbative $h_R$, this cannot be satisfied. For $J\approx 1~\mathrm{eV}$, the right-hand side is (0.14nm)$^{-1}$. On the other hand, $k_\perp^{\rm max}$ is bounded by the order of the inverse lattice parameter, which is around (0.3nm)$^{-1}$. Therefore, $h_R$ should be greater than one to satisfy Eq.~(\ref{Eq(A):Bound state condition}), which is beyond a perturbative regime.

\end{appendix}

\end{document}